\documentstyle[a4,12pt,psfig]{article}

\newcommand{\Gslash}[1]{ \parbox[b]{1em}{$#1$} \hspace{-0.95em}
                         \parbox[b]{0.55em}{ \raisebox{0.2ex}{$/$}}}
\newcommand{\Sslash}[1]{ \parbox[b]{0.6em}{$#1$} \hspace{-0.55em}
                         \parbox[b]{0.55em}{ \raisebox{-0.2ex}{$/$}}}
\newcommand{\beq}{\begin{equation}}
\newcommand{\eeq}{\end{equation}}
\newcommand{\beqa}{\begin{eqnarray}}
\newcommand{\eeqa}{\end{eqnarray}}

\newcommand{\M}{{\cal M}}

\newcommand{\pvec}{{\bf p}}

\newcommand{\qvec}{{\bf q}}
\newcommand{\epsvec}{{\bf \varepsilon}}
\newcommand{\evec}{{\bf e}}
\newcommand{\lvec}{{\bf \ell}}

\newcommand{\epsla}{\Sslash{\varepsilon}}

\newcommand{\psla}{\Sslash{p}}

\newcommand{\gsla}{\Gslash{g}}

\newcommand{\as}{\alpha_s}

\begin{document}
\begin{titlepage}
\title{\bf \Large
\vspace{2cm}
Quarkonium Production through Hard Comover Rescattering in 
Polarized and Unpolarized $pp$ Scattering.}
\author{Martin Maul
\\ \\ \\ \small
Department of Theoretical Physics, Lund University,
S\"olvegatan 14A, S - 223 62 Lund, Sweden}
\maketitle
\thispagestyle{empty}
\vspace{-4.5truein}
\begin{flushright}
{\bf LU TP 00-33}\\
{\bf hep-ph/0009279}\\
{\bf \date{\today}}
\end{flushright}
\vspace{4.0truein}
\begin{abstract}
In this paper hadroproduction of charmonium states in
polarized $pp$ collisions is discussed. A thermal picture for the gluonic
cloud of comovers is given making contact between the formalism
and the measured unpolarized cross sections. The experimentally
observed non-polarization of the final $J/\psi$ states leads to
the consequence that no correlations between the initial proton spin
and the final charmonium spin should be existent. Hence the single spin
asymmetries vanish to leading order in that model.
\end{abstract}
\vskip4cm\par
{{\bf PACS}: 12.39.Hg, 29.27.Hj, 29.25.Pj}
\end{titlepage}
\section{Introduction}
\label{secintro}
Charmonium production in polarized $pp$ scattering at RHIC is one 
of the key experiments to pin down the polarized gluon distribution
amplitude \cite{Ramsey:2000wx}. 
The standard formalism in which heavy quarkonium production
in hadron hadron collisions is  described is still the 
Color Octet Mechanism (COM)  \cite{Jaffe:1999ww}. However,
it is known that this mechanism fails to describe the experimentally
observed non-polarization of the final $J/\psi$ and $\psi'$
 meson \cite{Affolder:2000nn}
and, furthermore,
it predicts a $\chi_{c1}/\chi_{c2}$ production ratio which is far 
too low \cite{Beneke:1996tk}. 
Recently, it has been shown that these problems can be 
cured by assuming that the charmonium formation happens through
rescattering with a gluon cloud of hard comovers
\cite{Hoyer:1999ha,Marchal:2000wd}:
The two colliding hadrons form through
the self interacting gluon field a gluonic
medium in which the heavy quarkonium formation is directed
by hard rescattering processes. This means a crucial qualitative
difference as to electroproduction where one of the collision partners
is a lepton and cannot participate in strong interactions.
As a deeper understanding of the
heavy quarkonium formation mechanism in polarized $pp$ scattering
will be very important for the extraction of the polarized
gluon density, we want to  investigate what consequences this theory 
implies for the charmonium production in polarized $pp$  scattering. 
In addition to the rescattering picture developed in 
\cite{Hoyer:1999ha,Marchal:2000wd} we will present a thermal description
of the comover cloud.
\newline
\newline
Double spin asymmetries in charmonium hadroproduction have been calculated
in the framework of the COM formalism, which is based on a systematic
non-relativistic velocity expansion \cite{Bodwin:1995jh},
 for the prospective HERA-$\vec N$ 
experiment 
\cite{Teryaev:1997sr,Nowak:1997ic,Korotkov:1997sa,Korotkov:1999in,Nowak:1998jy}.
In the framework of the Color Singlet Model (CSM) 
\cite{Berger:1981ni,Baier:1981uk} double spin asymmetries
in $J/\psi$ production have been studied in 
\cite{Morii:1994ja}. The CSM has been shown to be incompatible
with the absolute size of the unpolarized $J/\psi$ cross section
\cite{Beneke:1996tk}. Attempts to cure
this failure by the
$k_\perp$-factorization approach within the CSM have
not led to satisfying results \cite{Yuan:2000cp,Hagler:2000eu}. Therefore
one is looking for a suitable combination of the CSM and the COM
formalism in the $k_\perp$-factorization approach which may release
the polarization problem of the final $J/\psi$ and $\psi'$ \cite{Yuan:2000qe}. 
A discussion
for the double spin asymmetry for RHIC energies in
terms of the CSM  formalism can be found in 
\cite{Morii:1996xb}. 
Higher order velocity corrections in
the polarized case in terms of the COM formalism have
been discussed in \cite{Gupta:1998nw,Gupta:1998ie,Gupta:1997nj}.
It is the aim of this paper to add to this discussion the possible
insights the comover rescattering picture can offer as to polarized
hadroproduction of charmonium states.
\newline
\newline
In  Secs.~\ref{secswave} and \ref{secpwave} we will write
down the total cross section for polarized S-wave and P-wave
charmonium production. For the gluonic comovers we will
develop a thermal description as a boson gas and explain
the consequences for the polarized partonic cross sections.
in Secs.~\ref{secfitmass} and \ref{secfitvol} we will fit
the parameters of the theory (volume of the comover cloud, the
energy the charm quarks carry on the average, and the expansion
parameter $\rho$ of the charmonium system)
to the numerical values of the measured unpolarized
charmonium cross sections. We will discuss the implications
on the energy transfer from the colliding particles to the cloud
and from the comover cloud to the charmonium system. We give also
a picture of the geometry of the cloud. Finally, in 
Sec.~\ref{secall} we will calculate the double spin asymmetries
for inclusive $J/\psi$, $\psi'$ and $\chi_{cJ},J=0,1,2$
production in the framework of the thermal description. 

\section{Cross section for S-wave quarkonium production}
\label{secswave}
\begin{figure}[tb]
\centerline{\psfig{figure=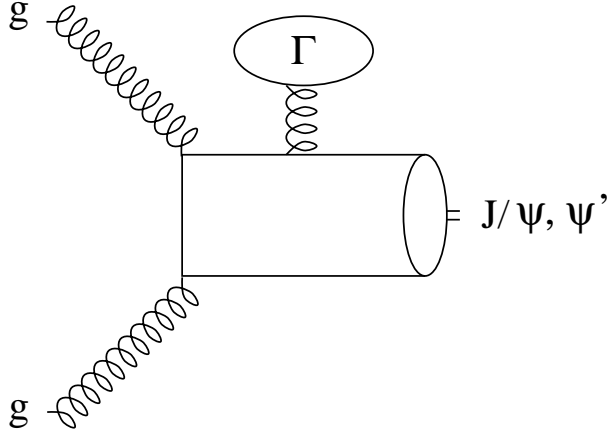,width=8cm}}
\caption{Typical diagram for the  S-wave charmonium production amplitude.
The formation of the $J/\psi$ mesons happens through hard interaction
with a gluonic cloud of comovers ($\Gamma$).}
\label{ggjpsi}
\end{figure}

The gg-fusion amplitude for $^3S_1$ quarkonium production in the presence 
of a background gluon field $\Gamma_\mu(\lvec)$ 
(see Fig.~\ref{ggjpsi} for a typical diagram) has been derived in 
\cite{Hoyer:1999ha,Marchal:2000wd}:
\beqa
\M(^3S_1,S_z) &=& 
\frac{1}{2} d^{abc}   \frac{g^3R_0}{\sqrt{6\pi m^3}}
\left\{ i\lambda_1 \delta_{\lambda_1}^{-\lambda_2} {\bf \Gamma}^c
(\lvec) \times \lvec \cdot \evec(S_z)^* \right. \nonumber \\
&&+ \left. \Gamma_0^c(\lvec) \left[ -\delta_{\lambda_1}^{-\lambda_2}
\ell^z \delta_{S_z}^0 + \evec(\lambda_1) \cdot \lvec
\delta_{S_z}^{\lambda_2} + \evec(\lambda_2) \cdot \lvec
\delta_{S_z}^{\lambda_1} \right] \right\} \;. 
\label{swaveampl}
\eeqa
%
%
%
%
Here $R_0$ is the value of the S-wave quarkonium wave function at the origin.
$2m$ is the energy of the 
two incoming gluons. To the accuracy of the first order
velocity expansion used here it is identical
to the energy of the charm quarks before and after the interaction
with the comover cloud and also identical to the mass
of the $J/\psi$ or $\psi'$  produced. Effectively, we will 
see later that we have to adjust the numerical value for $m$ to
experimental data of the unpolarized cross section.
$\lambda_1,\lambda_2$ 
are the helicities of the two incoming partonic gluons with polarization
$\evec(\lambda_i),i=1,2$. It has been pointed out in 
\cite{Hoyer:1999ha,Marchal:2000wd} that the fact that
the polarization of  $J/\psi$ and $\psi'$ is small and consistent
with zero, as can be seen from fixed target data 
\cite{Badier:1983dg,Biino:1987qu,Akerlof:1993ij,Gribushin:1996rt,Alexopoulos:1997yd,Heinrich:1991zm} and also from the new CDF data \cite{Affolder:2000nn},
leads to the condition that in the c.m. system of the two quarks the relation:
\beq
| \Gamma_0^c(\lvec)|^2 \ll 
\left| {\bf \Gamma}^c(\lvec)\times \lvec \right|^2/\lvec^2
\label{unpolcond}
\eeq 
must be fulfilled. $\Gamma_\mu^a = \frac{1}{\sqrt{N_c^2-1}}
\Gamma_\mu $ is the four-vector potential of the interacting
gluonic field. The philosophy of our  concept is that the quarkonium
production happens in a heat bath of essentially co-moving (real) gluons.
The polarization tensor of the real gluons can then be described by:
\begin{equation}
\Gamma_\mu(\lvec) \Gamma^*_{\mu'}(\lvec)
= -g_{\mu{\mu'}} + \frac{\ell_\mu n_{\mu'} + \ell_{\mu'} n_\mu}{\ell\cdot n}
- \frac{\ell_\mu \ell_{\mu'} n^2}{(\ell \cdot n)^2}\;.
\end{equation}
The introduction of a real gluon field means on the other hand that
the incoming charm quark from the gluon gluon fusion amplitude 
must be slightly off-shell. In the calculation this off-shellness
is not included. In fact, it can be regarded as a velocity correction
of order $v = 2l_0/m$ (see App.~\ref{virtreal})
which we will neglect in the calculations that follow. The other alternative,
i.e. to have a virtual field $\Gamma$ leads together with the requirement
$\Gamma_0 \approx 0$ to unphysical consequences as we show in  App.~\ref{virtreal},
whereas all requirements turn out to be natural for a real gluon field $\Gamma$.
For the $\psi'$ and $\chi_{c1}$ asymmetries discussed in Sec.~\ref{secall} all
model dependent parameters cancel and we coincide exactly with the
model for $\Gamma$ proposed in  \cite{Hoyer:1999ha,Marchal:2000wd}.
\newline \newline
In order to fulfill the condition that ensures the non-polarization
of quarkonium states Eq.~(\ref{unpolcond}), we parametrize
$n = (1,b\evec)$, with $\evec$ being a unit vector and $0\le b \ll 1$.
In fact, for $b=0$ we find $\Gamma_0 \equiv 0$. One can interpret
$b$ approximately as the relative velocity of the gluon comovers to the
c.m. system of the two incoming gluons
 and $\evec$ as the direction of this relative
movement as to the axis defined by the two incoming gluons which is 
the z-axis in our case. In the following we will refer to $b$ then
as displacement parameter.
\newline
\newline
The heat bath, in which the quarkonium production 
takes place has the temperature $T=1/\beta$ and is described by the 
Bose-Einstein statistics. This takes into account the multiple gluon
interaction inside the cloud.
Guided by this philosophy, we define the 
following three quantities:
\begin{eqnarray}
\Gamma^{(V)}_P &=& V \int \frac{d^3 \lvec}{2|\lvec|(2\pi)^3} 
\frac{|({\bf \Gamma}(\lvec) \times \lvec)\cdot \evec^*(P)|^2}
     {\exp(\beta (n\cdot \ell))-1}
\nonumber \\
&\approx& V \frac{\zeta(4)}{\pi^2\beta^4}
\nonumber \\
\nonumber \\
\Gamma^{(E)}_P &=& V \int \frac{d^3 \lvec}{2|\lvec|(2\pi)^3} 
\frac{|( \Gamma_0(\lvec) (\lvec\cdot \evec^*(P))|^2)}
     {\exp(\beta (n\cdot \ell))-1}
\nonumber \\
&\approx& 
V\frac{b^2\zeta(4)}{10\pi^2\beta^4}
\left[4 -\left(
 |\evec \cdot \evec(P)|^2 + |\evec \cdot \evec^*(P)|^2\right)
\right]
\nonumber \\
\nonumber \\
\Gamma^{({\rm int})}_{S_z,P} &=&  2 V \int \frac{d^3 \lvec}{2|\lvec|(2\pi)^3} 
\frac{{\rm Im}\left[ ({\bf \Gamma} (\lvec)\times \lvec)^* \cdot \evec(S_z) 
\Gamma_0(\lvec) (\evec(P)\cdot \lvec)\right]}
     {\exp(\beta (n\cdot \ell))-1}
\nonumber \\
&\approx& 
V\frac{b\zeta(4)}{\pi^2\beta^4}
{\rm Im} \left[ \left( \evec \times \evec(P)\right)\cdot \evec(S_z)\right]\;.
\label{thermal}
\end{eqnarray}
One should notice that with this form we reproduce the Stefan-Boltzmann
law, that the energy density represented by the field squared grows
$ \sim T^4$ with temperature.
The polarization $P$ can take the values $P=-1,0,+1$, and the polarization
vector $\evec(P)$, (and also  $\evec(S_z)$) is defined by:
\begin{equation}
\evec(P) = \left\{ \begin{array}{cc}
                   (-1,-i,0)/\sqrt{2} & {\rm if} \quad P = +1 \\
                   (\;\; 0, \;\;0,1)\quad \;\; & {\rm if} \quad P =\; \;0 \\
                   (+1,-i,0)/\sqrt{2} & {\rm if} \quad P = -1 
                   \end{array} \right. \;.
\end{equation}
We then can write down the following partonic cross section:
\beqa
\sigma_{\lambda_1\lambda_2 S_z} &=& 
\frac{2\pi}{2(2m)^3} \sum_{abc} \frac{1}{(N_c^2-1)^2}
 \left|\M(^3S_1,S_z)\right|^2
\nonumber \\
&=& \frac{5}{9}  \frac{\pi^3 \as^3 R_0^2}{(2m)^6}
\Bigg[ \delta_{\lambda_1}^{-\lambda_2} 
\left(\Gamma^{(V)}_{S_z} + \delta_{S_z}^0\Gamma^{(E)}_0 \right) 
 +\delta_{S_z}^{\lambda_1} \Gamma^{(E)}_{\lambda_2}
+
\delta_{S_z}^{\lambda_2} \Gamma^{(E)}_{\lambda_1}
+
2 \delta_{S_z}^{\lambda_2}\delta_{S_z}^{\lambda_1 } \Gamma^{(E)}_{S_z}
\nonumber \\
&&  \qquad  \qquad  \qquad
-\lambda_1 \delta_{\lambda_1}^{-\lambda_2} 
\left(
\delta_{S_z}^0  \Gamma^{({\rm int})}_{S_z,0}
-
\delta_{S_z}^{\lambda_1} \Gamma^{({\rm int})}_{S_z,\lambda_2}
-
\delta_{S_z}^{\lambda_2} \Gamma^{({\rm int})}_{S_z,\lambda_1}
\right)\Bigg]
\nonumber \\
&=& \frac{5}{9}  \frac{\pi^3 \as^3 R_0^2}{(2m)^6}
\Bigg[ \delta_{\lambda_1}^{-\lambda_2} 
       \left(\Gamma^{(V)}_{S_z} +  \Gamma^{(E)}_{-S_z}\right)
      + 4 \delta_{\lambda_1}^{\lambda_2} \delta^{\lambda_1}_{S_z}
        \Gamma^{(E)}_{S_z} 
      + \lambda_1 \delta_{\lambda_1}^{-\lambda_2} 
       \left(1-2\delta_{S_z}^0 \right)   \Gamma^{({\rm int})}_{S_z,-S_z}
\Bigg]\;.
\eeqa
The corresponding hadronic cross section is given by:
\beqa
\Sigma_{\Lambda_1\Lambda_2S_z} &=& 
\frac{d \sigma_{h \Lambda_1 \Lambda_2 S_z}}{dx_1 dx_2}
\nonumber \\
&=& \frac{1}{4}\sum_{\lambda_1\lambda_2} 
\left(G(x_1,(2m)^2) + \Lambda_1 \lambda_1 \Delta G(x_1,(2m)^2)\right)
\nonumber \\ && \times
\left(G(x_2,(2m)^2) + \Lambda_2 \lambda_2 \Delta G(x_2,(2m)^2)\right)
\sigma_{\lambda_1\lambda_2 S_z}
\delta \left(1 - \frac{(2m)^2}{Sx_1x_2}\right)\;.
\eeqa
Now one can isolate the various components:
\beqa
\Sigma_{00}(S_z) &=& \frac{1}{4}\left[
\left(\Sigma_{++S_z} + \Sigma_{--S_z} \right) +
\left(\Sigma_{+-S_z} + \Sigma_{-+S_z} \right)\right] 
\nonumber \\ &=&  \frac{1}{4}
\sum_{\lambda_1\lambda_2} 
 G(x_1,(2m)^2) G(x_2,(2m)^2)
\sigma_{\lambda_1\lambda_2 S_z}
\nonumber \\
&=&{\cal F}_S G(x_1,(2m)^2) G(x_2,(2m)^2) 
\left[ 
  \left( 1-\delta_{S_z}^0 \right) \Gamma_{S_z}^{(E)}
+\frac{1}{2}\left(\Gamma_{S_z}^{(V)} + \Gamma_{-S_z}^{(E)}\right) 
\right]
\nonumber \\ 
\nonumber \\ 
\Sigma_{LL}(S_z) &=&  \frac{1}{4}\left[
\left(\Sigma_{++S_z} + \Sigma_{--S_z} \right) -
\left(\Sigma_{+-S_z} + \Sigma_{-+S_z} \right)  \right]
\nonumber \\ &=&  \frac{1}{4}
\sum_{\lambda_1\lambda_2} 
\lambda_1\lambda_2\Delta G(x_1,(2m)^2)\Delta G(x_2,(2m)^2)
\sigma_{\lambda_1\lambda_2 S_z}
\nonumber \\ 
&=& {\cal F}_S \Delta G(x_1,(2m)^2) \Delta G(x_2,(2m)^2) 
\left[ 
  \left( 1-\delta_{S_z}^0 \right) \Gamma_{S_z}^{(E)}
-\frac{1}{2}\left(\Gamma_{S_z}^{(V)} + \Gamma_{-S_z}^{(E)}\right) 
\right]
\nonumber \\ 
\nonumber \\ 
\Sigma_{L0}(S_z) &=&  \frac{1}{4}\left[
\left(\Sigma_{++S_z} - \Sigma_{--S_z} \right) +
\left(\Sigma_{+-S_z} - \Sigma_{-+S_z} \right)\right] 
\nonumber \\ &=&  \frac{1}{4}
\sum_{\lambda_1\lambda_2} 
 \lambda_1  \Delta G(x_1,(2m)^2)  G(x_2,(2m)^2)
\sigma_{\lambda_1\lambda_2 S_z}
\nonumber \\ 
&=&{\cal F}_S \Delta G(x_1,(2m)^2)  G(x_2,(2m)^2) 
\left[
S_z  \Gamma_{S_z}^{(E)} 
+\frac{1}{2}\left(1-2\delta_{S_z}^0\right) \Gamma_{S_z,-S_z}^{(\rm int)} 
\right]
\nonumber \\ 
\nonumber \\ 
\Sigma_{0L}(S_z) &=&  \frac{1}{4}\left[
\left(\Sigma_{++S_z} - \Sigma_{--S_z} \right) -
\left(\Sigma_{+-S_z} - \Sigma_{-+S_z} \right) \right] 
\nonumber \\ &=&  \frac{1}{4}
\sum_{\lambda_1\lambda_2} 
 \lambda_2  \Delta G(x_2,(2m)^2)  G(x_1,(2m)^2)
\sigma_{\lambda_1\lambda_2 S_z}
\nonumber \\
&=& 
{\cal F}_S \Delta G(x_1,(2m)^2)  G(x_2,(2m)^2) 
\left[
S_z  \Gamma_{S_z}^{(E)} 
-\frac{1}{2}\left(1-2\delta_{S_z}^0\right) \Gamma_{S_z,-S_z}^{(\rm int)} 
\right]\;.
\label{asymmetries}
\eeqa
The common pre-factor is: 
\beq
{\cal F}_S((2m)^2) =
\frac{5}{9} \frac{\pi^3\as^3 R_0^2}
{(2m)^{6}} \delta\left(1 - \frac{(2m)^2}{Sx_1 x_2}\right)\;,
\eeq
where $(2m)$ is the energy of the two incoming gluons.
Eq.~(\ref{asymmetries}) is valid in general, even without any assumptions
about the thermal nature of the gluonic cloud, if we modify the definition
of $\Gamma^{(V,E,{\rm int})}$ accordingly. In case we apply our model
we find always $\Gamma_{S_z,-S_z}^{(\rm int)} = 0$. Then, 
the single spin asymmetries in our model are proportional to $S_z \Gamma^{(E)}_{S_z}$,
which means proportional $b^2$. As we know from the non-polarization of the final
$J/\psi$ and $\psi'$ meson that $b$ must be small and consistent with zero,
we consequently predict the absence of single spin asymmetries in charmonium
hadroproduction. Furthermore, if we can set $b$ to zero as the non-polarization
of the final $J/\psi$ and $\psi'$ suggests, we see that no initial spin - final spin
correlations are present and the double spin asymmetry reduces to:
\beq
\frac{\Sigma_{LL}(S_z)}{\Sigma_{00}(S_z)} \approx 
\frac{\Delta G(x_1,(2m)^2) \Delta G(x_2,(2m)^2)}{ G(x_1,(2m)^2) 
G(x_2,(2m)^2)}\;.
\label{dsa}
\eeq
In other words, the experimental fact that  
the $J/\psi$ and $\psi'$ are produced in a non-polarized mode assures
that the double spin asymmetry Eq.~(\ref{dsa}) is essentially
only dependent on the ratio of the unpolarized and polarized gluon
parton distributions, which is a very important statement as far
as the possibility is concerned to isolate the polarized gluon 
density from direct $J/\psi$ or $\psi'$ production data. 
\newline \newline
We can describe the gluon cloud of comovers with a Bose distribution of momenta
in the sense of a gluon plasma, in which the formation of the charmonium states
takes place. To get a quantitative model the following quantities have to be 
set:
\begin{itemize}
\item The temperature $T$ of the cloud which should lie above the phase transition of 
hadronic matter, i.e. larger than typically $200\;{\rm MeV}$. It should also be much
smaller than the typical charm mass of $1.5\;{\rm  GeV}$, otherwise the interaction
with the gluon cloud
would rather inhibit the charmonium production than catalyze it.
\item The volume $V$ of the cloud which should be large enough to comprise the $c\bar c$
      system.
\item The energy $(2m)$ of the two initial gluons.
\item The expansion parameter $\rho$ from the quarkonium wave function at the origin
      to the interaction point with the comovers.
\end{itemize}
From the non-polarization of the final $J/\psi$ and $\psi'$ we can already set the 
parameter $b=0$ for the following. We will do this analysis in Sec.~\ref{secfitmass} and
Sec.~\ref{secfitvol} after having collected the cross sections for the P-wave 
charmonium production in the next section.

\section{Cross section for P-wave charmonium production}
\label{secpwave}
\begin{figure}[tb]
\centerline{\psfig{figure=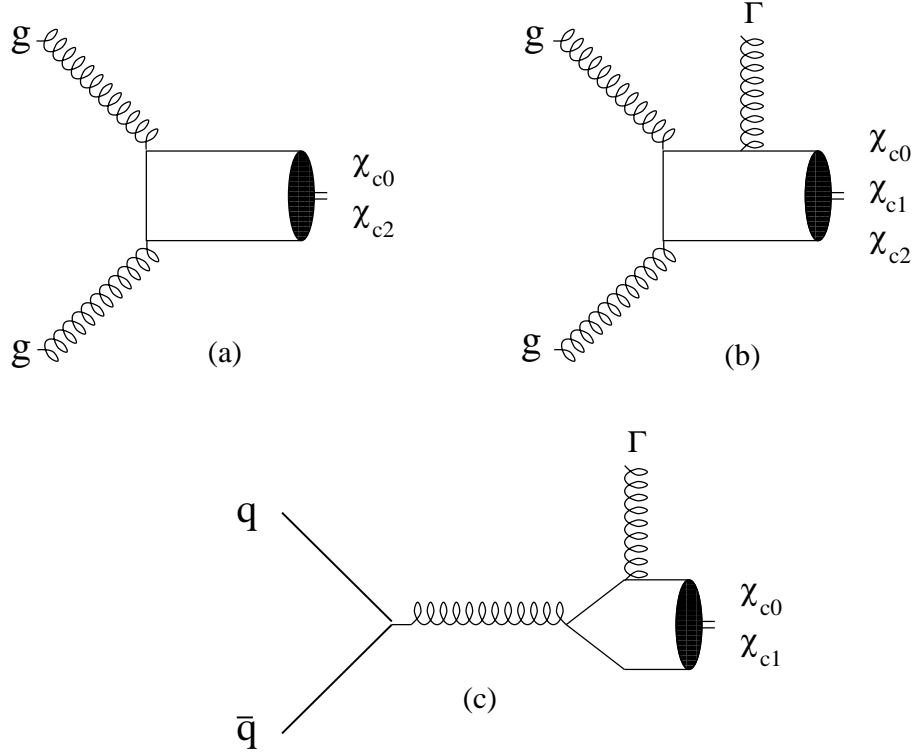,width=12cm}}
\caption{Typical diagrams for the  P-wave charmonium production: (a) CSM
contribution without comovers at ${\cal O}(\as^2)$;  contributions
with comover interaction at ${\cal O}(\as^3)$: gg fusion (b) and $q\bar q$
annihilation (c).}
\label{pwave}
\end{figure}

P-wave charmonium production for the mesons $\chi_{c0}$ and
$\chi_{c2}$ can occur via two gluons without any contribution
from comovers, see Fig.~\ref{pwave}(a). This corresponds
then to the contribution being calculated from
the Color Singlet Model (CSM). The comover contribution for
the production of $\chi_{cJ}, J=0,1,2$ mesons 
becomes important only at the ${\cal O}(\as^3)$ 
level, see Fig.~\ref{pwave}(b,c),
where the comover contribution is supposed to be dominant over 
all other CSM contribution of the same order in $\as$.
The ${\cal O}(\as^2)$ CSM partonic cross sections are given by: 
\beqa
\sigma^{(a)}_{\lambda_1 \lambda_2}(^3P_0,0)
&=& \frac{24\pi^2 \as^2 |R_1'|^2}{(2m)^7}  \delta_{\lambda_1}^{-\lambda_2} 
\nonumber \\
\sigma^{(a)}_{\lambda_1 \lambda_2}(^3P_1,J_z)
&=& 0
\nonumber \\
\sigma^{(a)}_{\lambda_1 \lambda_2}(^3P_2,J_z)
&=& \frac{32\pi^2 \as^2 |R_1'|^2}{(2m)^7}  \delta_{\lambda_1}^{\lambda_2} 
\delta_{J_z/2}^{\lambda_1} \;.
\eeqa
We show the derivation of these cross sections in App.~\ref{app1} to make
also sure for the constants needed to make contact between the thermal amplitudes
and the physical cross sections.
Averaging over the gluon helicities $\lambda_1$ and $\lambda_2$ and
summing over all possible final states $J_z$ we reproduce the unpolarized
partonic cross sections as given in 
\cite{Baier:1983va,Schuler:1994hy,Beneke:1996tk}.
%
%
%
%
At the level of three gluons, i.e. ${\cal O}(\as^3)$, we can again
use the formulas derived from hard comover scattering.
This time, however, it is 
hot possible to include a finite displacement $b$ as it would
require a further expansion of the gluon fusion amplitude, taking into
account the Lorentz transformation from the gluon gluon c.m.~to 
the final quark quark c.m.~\cite{Hoyer:1999ha}.
Such an expansion would destroy
the simple structure of the theory developed so far and therefore
we leave it out here. Furthermore, as data indicate from the
non-polarization of the final $J/\psi$ and $\psi'$ state should 
$b$ be small and consistent with zero. With this assumption
the P-wave amplitude reads \cite{Hoyer:1999ha}:
\beqa
\M^{(b)}(^3P_J,J_z) &=& \frac{\rho}{2} f^{abc} \frac{g^3R_1'/m}{\sqrt{6\pi
m^3}} i\lambda_1 \delta_{\lambda_1}^{-\lambda_2}
\nonumber \\ &&\times \left\{
\begin{array}{cc}
\left[ {\bf \Gamma}^c(\lvec) \times \lvec \right]^z & (J=0) \\
-i\sqrt{\frac{3}{2}} \left[ \evec^*(J_z) \times \left(
{\bf \Gamma}^c(\lvec)\times \lvec \right) \right]^z & (J=1) \\
-\sqrt{3} e_{3i}^*(J_z) \left[ {\bf\Gamma}^c(\lvec) \times \lvec
\right]^i & (J=2)
\end{array} \right. \;.
\label{3pjampl}
\eeqa
We define now additionally:
\beqa
\Gamma^{(V)}_{J=0,0} &=& \Gamma^{(V)}_{0}
\nonumber \\
\Gamma^{(V)}_{J=1,J_z} &=& V \int \frac{d^3 \lvec}{2|\lvec|(2\pi)^3} 
\frac{3}{2}
\frac{\left| \left[ \evec^*(J_z) \times \left(
{\bf \Gamma}(\lvec)\times \lvec \right) \right]^z\right|^2 }
{\exp(\beta (n\cdot \ell))-1}
= V \frac{3}{2}\frac{\zeta(4)}{\pi^2\beta^4}
\left( 1 - \delta_{J_z}^0\right)
\nonumber \\
\Gamma^{(V)}_{J=2,J_z} &=& V \int \frac{d^3 \lvec}{2|\lvec|(2\pi)^3} 
3 
\frac{\left| e_{3i}^*(J_z) \left[ {\bf\Gamma}(\lvec) \times \lvec
\right]^i \right|^2}
{\exp(\beta (n\cdot \ell))-1}
= V \frac{3}{2}\frac{\zeta(4)}{\pi^2\beta^4}
\left( \frac{4}{3} \delta_{J_z}^0 + \delta_{J_z}^{\pm 1}\right)\;.
\eeqa
The explicit form of the tensor  $e_{3i}^*(J_z)$ necessary for the calculation
performed here can be found in \cite{Cho:1996ce}. We find then for the partonic
cross section of the contribution from the diagrams of type Fig.~\ref{pwave}(b):
\beq
\sigma_{\lambda_1\lambda_2}^{(b)}(^3P_J,J_z) 
= 4 \rho^2 \frac{\pi^3\as^3 |R_1'|^2}{(2m)^8}
\delta_{\lambda_1}^{-\lambda_2} \Gamma^{(V)}_{J,J_z} \;.
\eeq
This formula states that there is no correlation between the helicities 
$\lambda_1,\lambda_2$ and the 
total angular momentum $J,J_z$. 
P wave quarkonium production can also occur through $q\bar q$ 
annihilation, see Fig.~\ref{pwave}(c). 
Using the formulas given in \cite{Marchal:2000wd},
the amplitude for this case is given by:
\beqa
{\cal M}^{(c)}_{ija}(^3P_J,J_z) &=& 
(2\lambda_2)\frac{\sqrt{1+|\lambda_1+\lambda_2|}}{(2m)^2}
 \frac{g^3}{2} \frac{\sqrt{3} R_1'}{\sqrt{2\pi m^3}} T_{ij}^a
\nonumber \\ && \qquad  \times 
\left\{ 
\begin{array}{cc} 
\evec(\lambda_1+\lambda_2) 
\cdot \left(2m\Gamma^a_0 \lvec + \ell^2 {\bf \Gamma^a} \right) 
  & (J=0) \\
i \sqrt{\frac{2}{3}}(\evec^*(J_z)\times \evec(\lambda_1 + \lambda_2))
 \left(2m\Gamma^a_0 \lvec + \ell^2 {\bf \Gamma^a} \right)      
  & (J=1) \\
0 & (J=2) \end{array} \right.\;.
\eeqa  
Here $\lambda_1$, $\lambda_2$ are the helicities of the incoming quark and antiquark, respectively.
$i,j$ are their color indices and $a$ is the color index of the gluon from the heat bath.
Then the partonic cross section reads:
\beqa
\sigma_{\lambda_1\lambda_2 J_z}^{(c)}(^3P_J,J_z) &=&
\frac{2\pi}{2(2m)^3} \sum_{ija} \frac{1}{N_c^2}
 |{\cal M}^{(c)}_{ija}(^3P_J,J_z)|^2
\nonumber \\ &=&
\frac{32\pi^3 \alpha_s^3 |R_1'|^2}{3(2m)^{10}}
\Gamma_{JJ_z}^{(T)} (\lambda_1 + \lambda_2)\;.
\eeqa  
If we refer to our model, discussed in the previous section and neglect 
$\Gamma_0$ we obtain:
\beqa
\Gamma_{JJ_z}^{(T)} (\lambda_1 + \lambda_2)
&=&(1+|\lambda_1+\lambda_2|)V \frac{20\zeta(6)}{\pi^2 \beta^6}
\left\{ 
\begin{array}{cc}
1 & (J=0) \\
\frac{2}{3}\left(1-| \evec(\lambda_1+\lambda_2) \cdot \evec(J_z)|^2 \right)
& (J=1) \\
0 & (J=2)
\end{array} \right.\;.
\eeqa
The essential statement is that the contribution (c) contains a correlation
between the spin of the charmonium and the spin of the two incoming gluons.
On the other hand this contribution $(c)$ is suppressed by a factor
$T^2/(2m)^2$ versus the contribution $(b)$. As $T^2/(2m)^2\sim |\lvec|^2/(2m)^2$
we can again neglect this contribution to the accuracy of the calculation
as it is of the same order as the velocity corrections not taken into
account here.
\newline \newline
The formalism used to derive the partonic cross sections is NRQCD to 
leading order. In principle to the order of accuracy of
this approximation the masses of all charmonium mesons 
are the same , 
i.e. $M_{J/\psi}$ = $M_{\psi'}$ = $M_{\chi_{cJ},J=0,1,2}$ = $2m_c$
\cite{Beneke:1996tk}. In the same order of accuracy we can also identify
$m =m_c$ in all the formulas above. One has to keep in mind that in
such a crude approximation the contribution of velocity corrections
may be quite substantial. Unfortunately, the inclusion of velocity corrections
will destroy the simple structure of the relations derived in
\cite{Hoyer:1999ha,Marchal:2000wd} and is therefore out of the scope of
the discussion here. To face this problem we will fit 
as a pragmatic ansatz $m$ for the different
charmonium states involved in a way that most experimental facts are
reproduced. $m$ plays then the role of the average gluon energy involved
in the production of the charmonium state under consideration.
\section{The choice of the quark-energy $m$}
\label{secfitmass}
The starting point of the consideration is the direct $J/\psi$ cross 
section. Here we can as a basis identify $2m= M_{J/\psi}$, because
the difference between the $M_{J/\psi} = 3.097\;{\rm GeV}$ and 
$2m_c = 3\;{\rm GeV}$ is small:
\beqa
\sigma_{\rm dir.}(J/\psi) 
&=&
\frac{V}{2} \int dx_1 dx_2 
{{\cal F}_S(M_{J/\psi}^2)  G(x_1,M_{J/\psi}^2) G(x_2,M_{J/\psi}^2)} 
\frac{\zeta(4)}{ \pi^2\beta^4}
\left[ 3 +\frac{b^2}{5}(11+2e_z^2)\right]
\nonumber \\
&=& \int d x_1 d x_2\sum_{S_z} \Sigma_{00}^{(J/\psi)}(S_z)\;.
\label{jpsidir}
\eeqa
To the order of accuracy of expansion we could apply the same arguments
also to the direct $\psi'$ production. However, such an approximation, which
would be in accordance of the standard COM velocity expansion, where 
 all charmonium masses are identical $M_{J/\psi}$= $M_\psi'$ = $M_{\chi cJ}$
= $2m_c$, is in practice very unsuitable as the masses enter in
high powers in the partonic cross section. We will therefore use
an effective mass for the direct $\psi'$ production:
\beqa
\sigma_{\rm dir.}(\psi') 
&=&
\frac{V}{2} \int dx_1 dx_2 
{{\cal F}_S(M_{\psi'\rm eff}^2)  G(x_1,M_{\psi' \rm eff}^2) G(x_2,M_{\psi'\rm eff }^2)} 
\frac{\zeta(4)}{ \pi^2\beta^4}
\left[ 3 +\frac{b^2}{5}(11+2e_z^2)\right]\;.
\eeqa
We can then fit $M_{\psi' {\rm eff}}$ \cite{Antoniazzi:1993iv} to the measured ratio 
$\sigma(\psi')_{\rm dir.}/\sigma(J/\psi)_{\rm dir.}$ given in \cite{Antoniazzi:1993iv}.
It has to be noticed that this ratio is completely independent of the cloud parameters.
The result of the fit is:
\begin{equation}
M_{\psi'{\rm eff}} =   3.4228 \; {\rm GeV} \;,
\end{equation}
which is a bit smaller than the real $\psi'$ mass of $M_{\psi'} =   3.6860 \;{\rm GeV}$. It
indicates in our language that there is a net transfer of energy from the gluon cloud into
the quarkonium system through hard comover rescattering or to stay in a thermal picture that
there is a transfer of energy from the hotter gluon cloud to the colder charmonium system.
The result of the fit can be seen in Tab.~\ref{psipdirtab}. \newline\newline
\begin{table}
\begin{center}
\begin{tabular}{|c||c|c|c|}
\hline
&&&\\
$\sigma(\psi')_{\rm dir.}/\sigma(J/\psi)_{\rm dir.}$ & GRV & CTEQ5L & experiment E705 \\
&&&\\
\hline
\hline
&&& \\
$p A$    &  0.23 & 0.20 & 0.21 $\pm$ 0.05 \\
$\pi A$  &  0.23 & 0.21 & 0.23 $\pm$ 0.05 \\
&&& \\
\hline
\end{tabular}
\end{center}
\caption{Fit results using  the best value $M_{\psi' {\rm eff}} =   3.4228 \; {\rm GeV}$
in comparison with experimental data from E705  \cite{Antoniazzi:1993iv}. The $p A$ experiment
was done at $E=300\;{\rm GeV}$ and the $\pi A$ experiment at $E=185\;{\rm GeV}$.
For the fit we get $\chi^2 = 0.08$ per degree of freedom.}
\label{psipdirtab}
\end{table}
We can now pay attention to the $\chi_{cJ}$ masses. As no comovers enter in the 
CSM contribution we should use in this case the original $\chi_{cJ}$ masses,
i.e.: 
\begin{equation}
\sigma^{\rm (CSM)}(\chi_{cJ}) =
\frac{1}{4}
\sum_{\lambda_1\lambda_2  J_z} \int dx_1 dx_2
G(x_1,M_{\chi_{cJ}}^2)G(x_2,M_{\chi_{cJ}}^2)
\sigma^{(a)}_{\lambda_1 \lambda_2}(^3P_J,J_z)
\delta \left( 1-\frac{M_{cJ}^2}{x_1 x_2 S}\right)\;.
\nonumber \\
\end{equation}
For the contribution resulting from comover rescattering we have to fit two
parameters, first the effective $\chi_{cJ}$ masses and then also the expansion
parameter $\rho$. We will proceed as follows. We take for all three  $\chi_{cJ}$
the same effective $\chi_{cJ}$ mass in the spirit that the effective mass
should be lowered proportional to what was the case for the $\psi'$ particle:
\begin{equation}
M_{\chi_{cJ} \rm eff} = \frac{1}{3} \left(\sum_{J=0}^2 M_{\chi_{cJ}} \right) 
\frac{M_{\psi'\rm eff}}{M_{\psi'}} = 3.2451\;{\rm GeV}\;. 
\end{equation}
For the contribution resulting from comover rescattering we can make a fit 
of the expansion parameter $\rho$ to
the E705 data by considering the reduced $\chi$ fraction, which is defined in a
way that  it is independent of the gluon cloud parameters in our approach:
\begin{eqnarray}
\sigma(J/\psi)_{\rm s-wave} &=& \sigma(J/\psi)_{\rm dir} + \sigma(\psi')Br(\psi' \to J/\psi) \nonumber \\
\sigma(J/\psi)_{\rm p-wave}^{(\rm CSM)} &=& 
\sum_J \sigma(\chi_{cJ})^{(\rm CSM)} Br(\chi_{cJ} \to J/\psi) \nonumber \\
\sigma(J/\psi)_{\rm p-wave}^{(\rm comovers)} &=& 
\sum_J \sigma(\chi_{cJ})^{(\rm comovers)} Br(\chi_{cJ} \to J/\psi) \nonumber \\
\left(\chi-{\rm frac}\right)_{\rm red} &=& 
\frac{\sigma(J/\psi)_{ \rm incl} - \sigma(J/\psi)_{\rm p-wave}^{(\rm CSM)} -\sigma(J/\psi)_{\rm s-wave}}
     {\sigma(J/\psi)_{ \rm incl} - \sigma(J/\psi)_{\rm p-wave}^{(\rm CSM)}}
\nonumber \\
&=& \frac{\sigma(J/\psi)_{\rm p-wave}^{(\rm comovers)}}
     {\sigma(J/\psi)_{\rm p-wave}^{(\rm comovers)}+\sigma(J/\psi)_{\rm s-wave} }\;.
\nonumber \\
\end{eqnarray}
In the framework of our thermal description this ratio is independent of the gluon cloud parameters.
Unfortunately, the reduced $\chi$ fraction is not directly measured which brings in an extra
dependence on the parton distributions used. The fit to the reduced $\chi$-fraction yields:
\begin{equation}
\rho  = 4.40 \;.
\end{equation}
This value is a bit larger than the value used in Ref.~\cite{Hoyer:1999ha}, where the
effective mass $M_{\chi_{cJ}\rm eff}$ was set to be $2 m_c$.
This points to the problem that the relative big
expansion parameter means that it may be inconsistent to consider only the
wave function of the quarkonium system at the origin. Besides the substantial
velocity corrections this is the second indication that the NRQCD approach in
general is not a suitable description of the problem.
In fact, future analysis
will have to find ways to go beyond the NRQCD approach which we have followed
here to be compatible with the standard literature of the field.
For the details of the fit using $\rho  = 4.40$ and 
$M_{\chi_{cJ} \rm eff}= 3.2451\;{\rm GeV}$   see Tab.~\ref{chifrac}.
\begin{table}
\begin{center}
\begin{tabular}{|c||c|c|c|c|}
\hline  &  \multicolumn{2}{c|}{ }      &\multicolumn{2}{c|}{ } \\  
        &  \multicolumn{2}{c|}{ $p A$} &\multicolumn{2}{c|}{ $\pi A$} \\
$\left(\chi-{\rm frac}\right)_{\rm red}$ 
         &  \multicolumn{2}{c|}{ }     &\multicolumn{2}{c|}{ } \\
&  experiment & fit & experiment & fit \\
&&&& \\ \hline \hline
&&&& \\
GRV    & 0.29  $\pm$ 0.04 & 0.34  & 0.36  $\pm$ 0.03  & 0.34 \\
CTEQ5L & 0.30  $\pm$ 0.04 & 0.33  & 0.36  $\pm$ 0.03  & 0.34 \\
&&&& \\
\hline
\end{tabular}
\end{center}
\caption{ Fit results using  the best value $\rho=4.40$ with 
 $M_{\chi_{cJ}' {\rm eff}} =    3.2451 \; {\rm GeV}$
in comparison with the values extracted from E705 experiment \cite{Antoniazzi:1993iv}. 
The $p A$ experiment was done at $E=300\;{\rm GeV}$ and the $\pi A$ experiment at $E=185\;{\rm GeV}$.
For the fit we get $\chi^2=  0.75 $ per degree of freedom.}
\label{chifrac}
\end{table}

\section{Determination of the other parameters of the theory}
\label{secfitvol}
\begin{figure}[tb]
\centerline{\psfig{figure=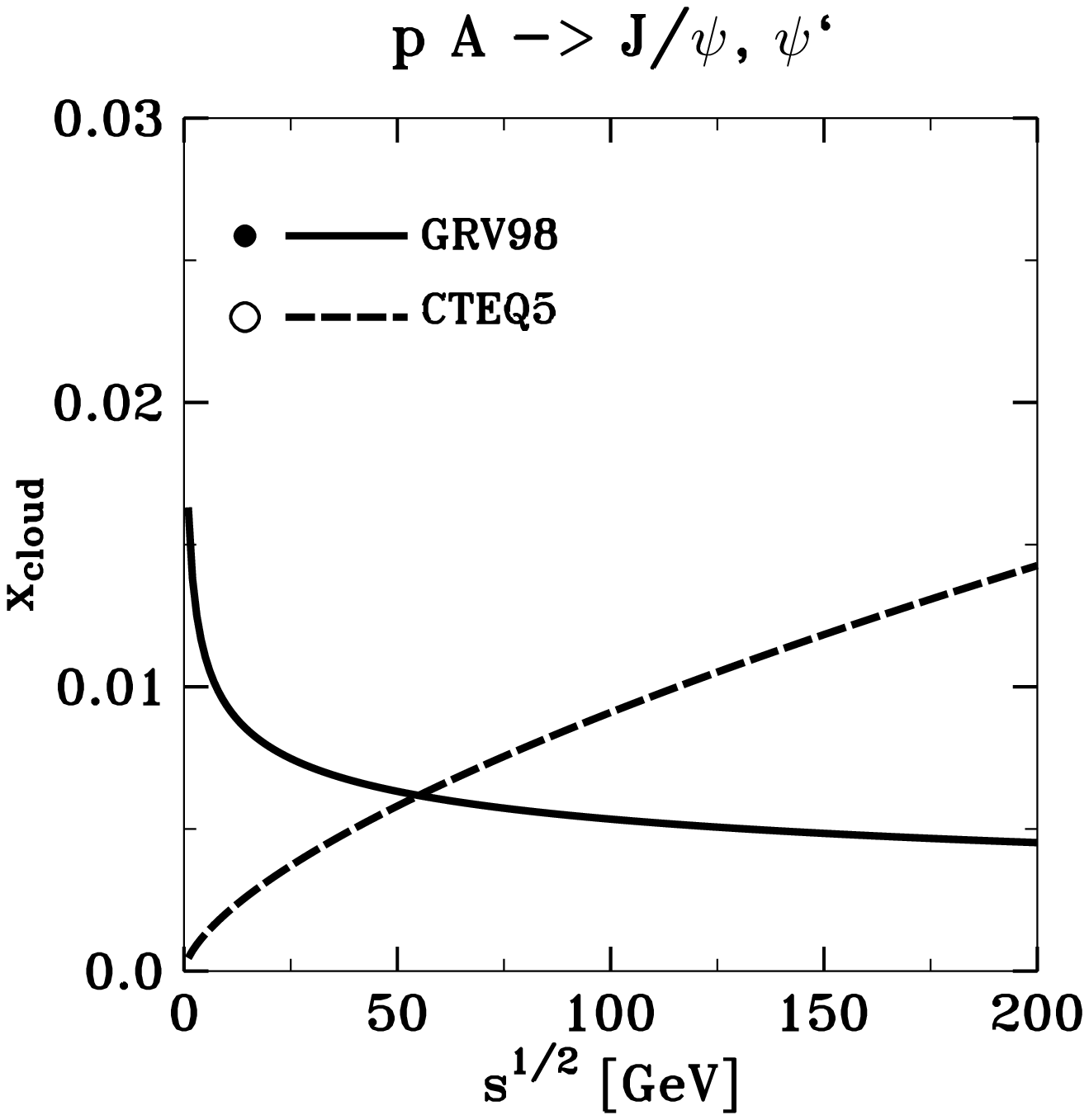,width=7cm}
            \psfig{figure=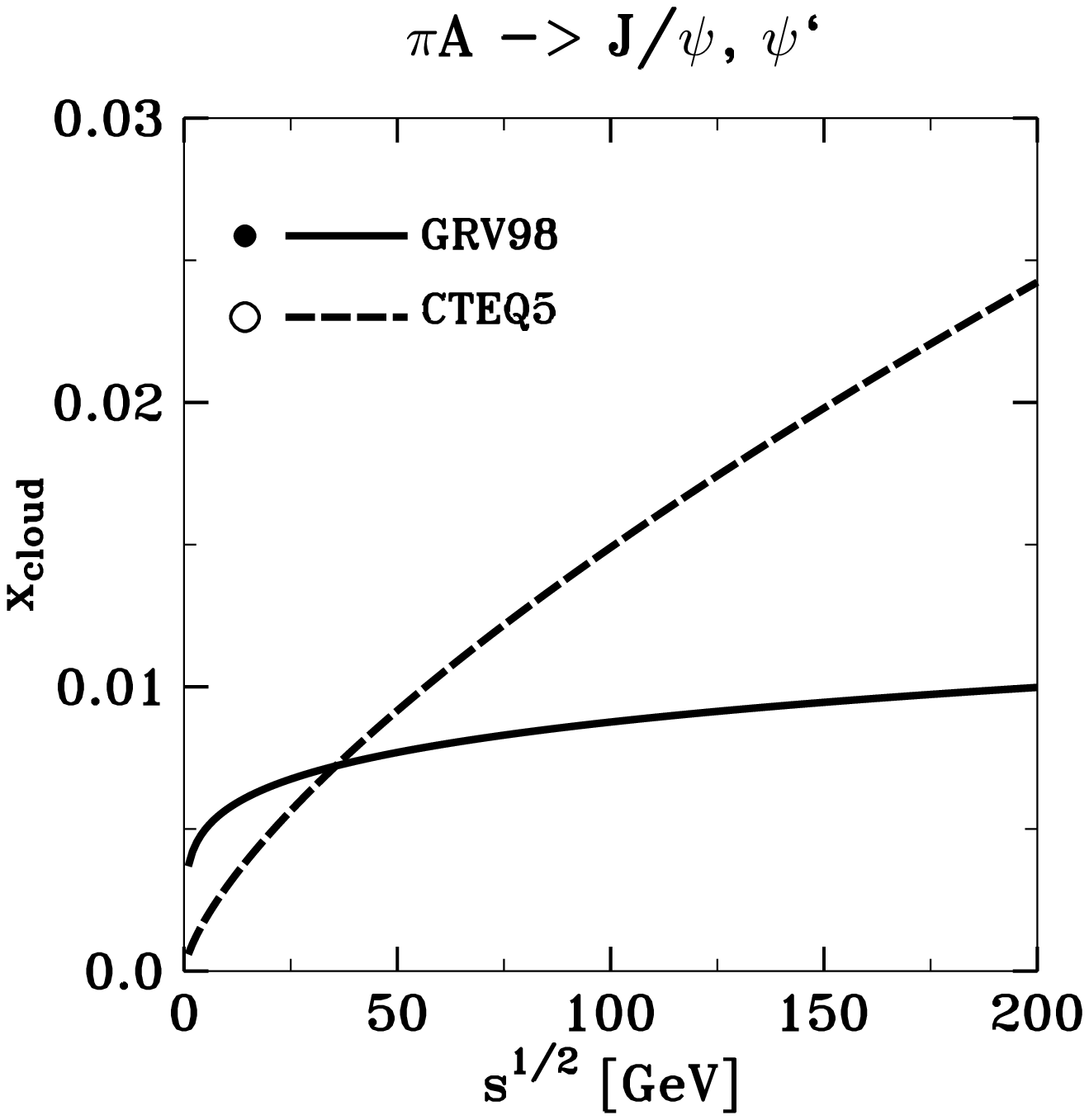,width=7cm}} 
\caption{The fraction of energy $x_{\rm cloud}$ transferred from the
projectiles to the gluon cloud in $pA$ collisions (right) and $\pi A$ 
collisions (left).}
\label{xcloud}
\end{figure}

The two parameters that are left undetermined so far are the active
Volume $V$ of the gluon cloud and its temperature $T$. From the
design of the theory the temperature is limited within tight bounds.
It must be well above $\Lambda_{\rm QCD}$ in order to make the 
interaction hard and to justify the use of perturbation theory. On the
other hand it must be smaller than the mass of the charm quark $m_c$, otherwise
the comover interaction would rather destroy the charmonium system than
to catalyze it.  Now
we can make an ansatz taking a constant value $T=500\;{\rm  MeV}$ and
let us check now what consequences this has for the active volume $V$.
For this purpose we will fit $V$ to the data available for inclusive
$J/\psi$ and $\psi'$ production. The situation is simple in the case of
the $\psi'$ production because it is a purely direct process:
\beqa
\sigma_{\rm dir.}(\psi') 
&=&
V\int dx_1 dx_2 \frac{{\cal F}_S(M_{\psi'{\rm eff}}^2)}{2}  
G(x_1,M_{\psi' {\rm eff}}^2) 
G(x_2,M_{\psi' {\rm eff}}^2) 
\frac{\zeta(4)}{ \pi^2\beta^4}
\left[ 3 +\frac{b^2}{5}(11+2e_z^2)\right] \;.
\nonumber \\
\label{psipdir}
\eeqa
The case of $J/\psi$ production is more complicated because
to a considerable amount  $J/\psi$ mesons can be produced indirectly via the decay
of $\chi_{cJ},J=0,1,2$ mesons predominantly through photon emission.
$(\chi_{cJ} \to J/\psi + \gamma)$ . Therefore we have to write for the total inclusive
unpolarized cross section for $J/\psi$ hadroproduction:
\beqa
\sigma_{\rm incl.}(J/\psi) = \sigma_{\rm dir.}(J/\psi) 
+  \sum_J  \sigma(\chi_{cJ}) {\rm Br}( \chi_{cJ} \to J/\psi +X)
+  \sigma_{\rm dir.}(\psi') {\rm Br}( \psi' \to J/\psi +X)\;.
\eeqa        
The total $\chi_{cJ}$ cross section is then given by the contribution 
from the diagrams in  Fig.~\ref{pwave} (a) and (b):
\beqa
\sigma(\chi_{cJ}) &=&  \sigma(\chi_{cJ})^{\rm (CSM)}
+ \sigma(\chi_{cJ})^{\rm(comovers)}
\nonumber \\
\sigma(\chi_{cJ})^{\rm (CSM)} &=&
\frac{1}{4}
\sum_{\lambda_1\lambda_2  J_z} \int dx_1 dx_2
G(x_1,M_{\chi_{cJ}}^2)G(x_2,M_{\chi_{cJ}}^2)
\sigma^{(a)}_{\lambda_1 \lambda_2}(^3P_J,J_z)
\delta \left( 1-\frac{M_{\chi_{cJ}}^2}{x_1 x_2 S}\right)
\nonumber \\
\sigma(\chi_{cJ})^{\rm (comovers)}                  &=& 
V\int dx_1 dx_2 G(x_1, M_{\chi_{cJ} \rm eff}^2)G(x_2, M_{\chi_{cJ} \rm eff}^2)
\frac{{\cal F}_P(M_{\chi_{cJ} \rm eff}^2)}{2} (2J+1)
\frac{\zeta(4)}{\pi^2\beta^4}\;,
\eeqa
with the pre-factor:
\beq
{\cal F}_P( M_{\chi_{cJ} \rm eff}^2)
 = 4 \rho^2 \frac{\pi^3 \as^3 |R_1'|^2}{M_{\chi_{cJ} \rm eff}^8} 
\delta\left(1-\frac{M_{\chi_{cJ} \rm eff}^2}{x_1 x_2 S} \right)\;.
\eeq
Numerically, the following branching ratios are used  \cite{Caso:1998tx}:
\beqa
{\rm Br}(\chi_{c0} \to J/\psi + \gamma ) &=& (\;\; 6.6 \pm 1.8)\times
10^{-3} \nonumber \\
{\rm Br}(\chi_{c1} \to J/\psi + \gamma ) &=& (27.3 \pm 1.6)\% \nonumber \\
{\rm Br}(\chi_{c2} \to J/\psi + \gamma ) &=& (13.5 \pm 1.1)\% \nonumber \\
{\rm Br}(\psi' \to  J/\psi + X)  &=& (54.2\pm 3.0)\% \;.
\eeqa
So putting all components together we get for the total inclusive $J/\psi$
cross section:
\beqa
\sigma_{\rm incl.}(J/\psi)
&=&
\int dx_1 dx_2 \Bigg\{
\frac{V\zeta(4)}{2\pi^2 \beta^4} \Bigg[
3 \Bigg(
 {G(x_1,M_{J/\psi}^2)G(x_2,M_{J/\psi}^2)}
 {\cal F}_S(M_{J/\psi}^2) 
\nonumber \\
&& \qquad \qquad\qquad \qquad   +
 {G(x_1,M_{\psi' \rm eff}^2)G(x_2,M_{\psi' \rm eff}^2)}
 {\cal F}_S(M_{\psi' \rm eff}^2) {\rm Br}(\psi' \to J/\psi)\Bigg)
\nonumber \\ && \qquad 
+ \sum_J (2J+1)
 {G(x_1,M_{\chi_{cJ} \rm eff}^2)G(x_2,M_{\chi_{cJ} \rm eff}^2)}
 {\cal F}_P(M_{\chi_{cJ} \rm eff}^2) {\rm Br}(\chi_{cJ} \to J/\psi) \Bigg]
\nonumber \\
&& +
\sum_J \sigma^{\rm (CSM)}(\chi_{cJ} ){\rm Br}(\chi_{cJ} \to J/\psi)\Bigg\}\;.
\label{jpsiincl}
\eeqa
For the data of the total cross section 
$\pi+A \to J/\psi,\psi'$ and $p(\bar p) + A \to  J/\psi,\psi'$  we
refer basically to \cite{Schuler:1994hy} and add the more
recent values from \cite{Schub:1995pu,Koreshev:1996wd,
Alexopoulos:1996dt,Akerlof:1993ij,Alexandrov:1999ch,Alexopoulos:1997yd}.
The cross sections have been rescaled to give the value over the 
whole range of $x_F$ 
(The details as to this rescaling are explained later in this chapter).
We reproduce essentially the figures in \cite{Beneke:1996tk}, except
for the data point from \cite{Akerlof:1993ij} for the $\pi N \to \psi'$
cross section, which is displayed a factor 2 too small. For the gluon
parton distribution of the proton we use the two leading order sets from CTEQ5
\cite{Lai:2000wy} and GRV98  \cite{Gluck:1998xa}. For the gluon parton
distribution in the pion we use the leading order parameterization 
given in \cite{Gluck:1999xe} (GRS99). For $\alpha_s$ we use the one-loop formula:
\begin{equation}
\alpha_s(\mu^2) = \frac{4\pi}{\left(11-\frac{2}{3}n_f\right)
\ln\left({\mu^2}/{(\Lambda_{\rm QCD}^{(n_f)})^2}\right)}
; \quad {\rm using} \; \Lambda_{\rm QCD}^{(4)} = 200\;{\rm MeV}\;,
\end{equation}
which comes close to  the value used in the GRV and CTEQ5L (leading order) gluon
distribution. The scale $\mu^2$ is given by the only scale relevant for
the partonic subprocess of quarkonium production, i.e. the quarkonium mass,
so $\mu^2 = M_{J/\psi}^2,  M_{\psi'}^2$, etc. The quarkonium wave function
at the origin $R_0$ is determined to leading order by the decay to $e^+e^-$:
\beq
\Gamma(J/\psi,\psi'\to e^+e^-) = \frac{4 e_c^2 \alpha_{\rm em}^2 R_0^2}
                    {M_{J/\psi,\psi'}^2}\;.
\eeq
We take the values 
$\Gamma(J/\psi\to e^+e^-) = 5.2374$ keV,
$\Gamma(\psi'\to e^+e^-) =  2.3545$ keV,
$M_{J/\psi} = 3097$ MeV,
$M_{\psi'}  = 3686$ MeV, $\alpha_{\rm  em}=1/137$ \cite{Caso:1998tx}. 
$e_c = 2/3$ is the charm
quark charge quantum number. 
In order to fix $|R_1'|^2$ we could try to extract it from the decay $\chi_{cJ} \to \gamma \gamma$,
however the data basis here is not very conclusive  \cite{Caso:1998tx}, and, therefore, we take here
in accordance with \cite{Hoyer:1999ha} and \cite{Beneke:1996tk} the value 
resulting from the Buchm\"uller-Tye potential given in \cite{Eichten:1995ch}, i.e.:
\begin{equation}
|R_1'|^2 = 0.075 \;{\rm GeV}^5 \;. 
\end{equation}
It should be noticed that other potentials, also tabulated in \cite{Eichten:1995ch}
yield considerable larger values for $|R_1'|^2$, up to nearly a factor of $2$. This
means that the expansion parameter then will be reduced by a factor $\sqrt{2}$ which 
will not help as to the principle problem mentioned above.
\newline
\newline
\begin{figure}[tb]
\centerline{\psfig{figure=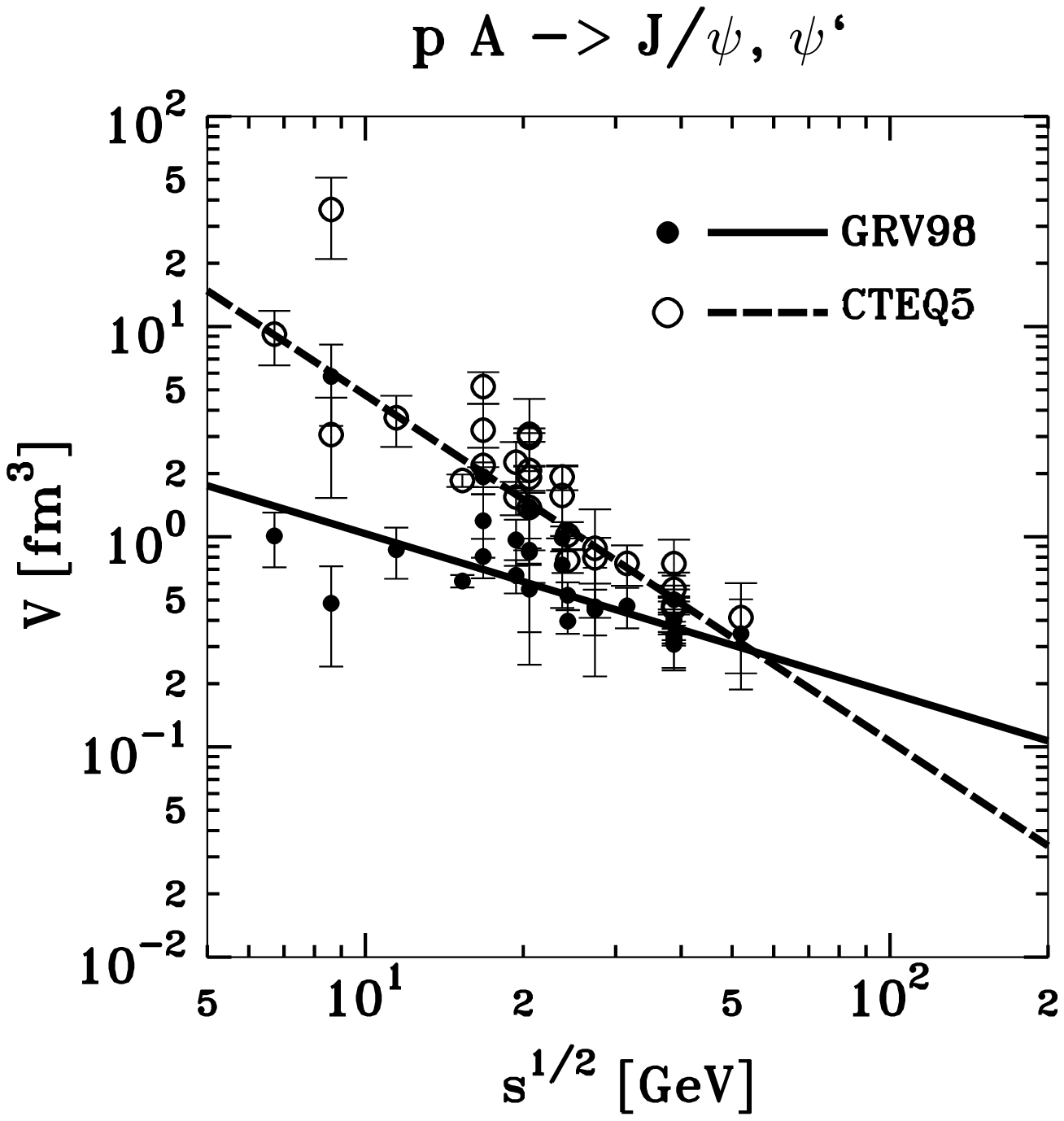,width=7cm}
            \psfig{figure=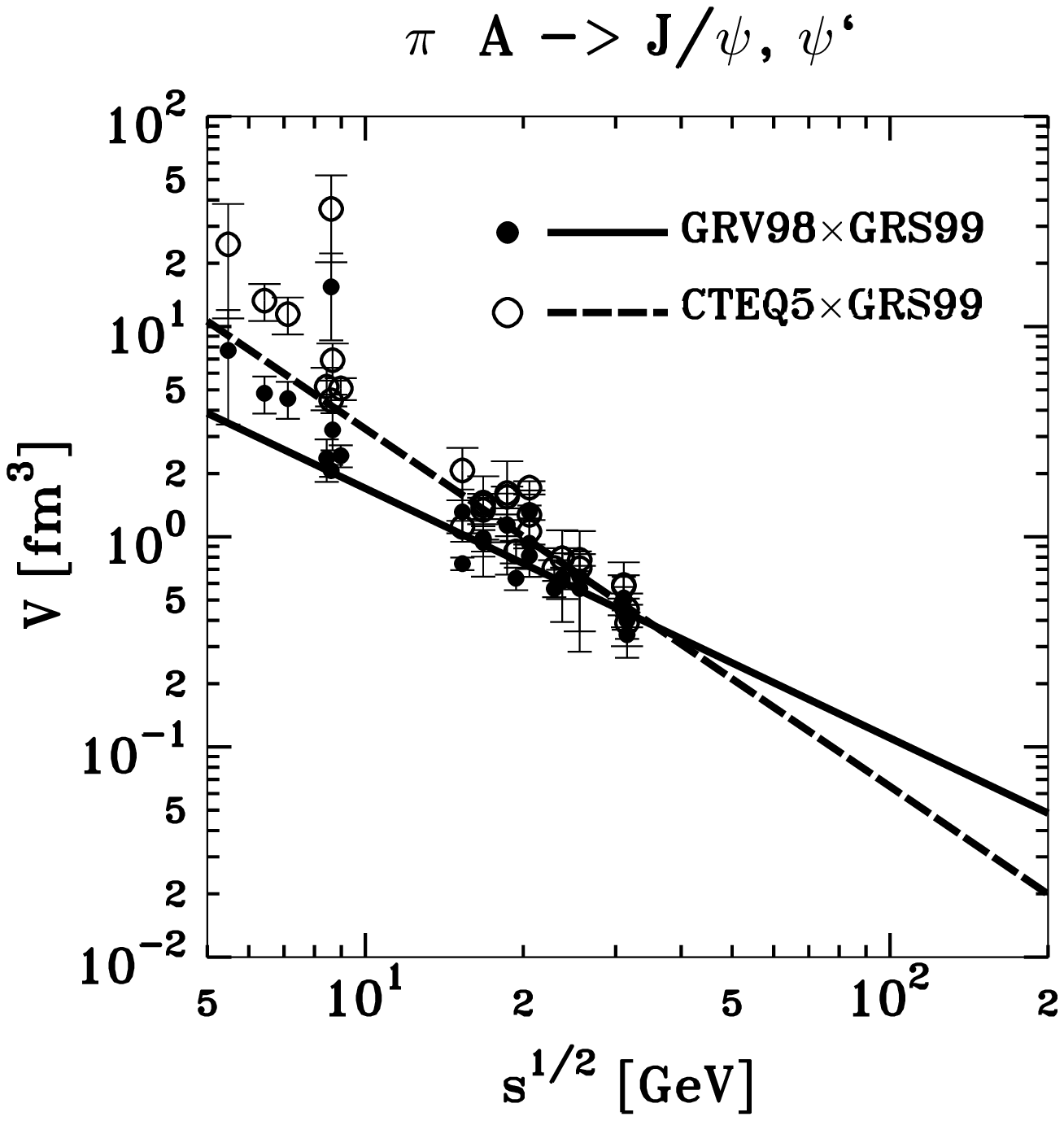,width=7cm}} 
\caption{The active volume of the gluon plasma fitted from $J/\psi$ and
$\psi'$ data  in $pA$ collisions (right) and $\pi A$ collisions
(left).}
\label{volume}
\end{figure}
%
%
Using Eqs.~(\ref{psipdir}) and (\ref{jpsiincl}) the active volume $V$ can be fitted to
the data available from $p A$ and $\pi A$ collisions. The result
is shown in Fig.~\ref{volume}. We assume a linear dependence in the
double logarithmic scale, i.e., a dependency of the  form 
$V = c \left(s/{\rm GeV}^2\right) ^p$.
The numerical results of the fit are shown in Tab.~\ref{volumefit}.
\newline
\newline
\begin{table}
\begin{center}
\begin{tabular}{|c||c|c|c|c|c|c|}
\hline 
 & \multicolumn{3}{c|}{ }   &  \multicolumn{3}{c|}{ }      \\
 & \multicolumn{3}{c|}{GRV} &  \multicolumn{3}{c|}{CTEQ5L} \\
 & \multicolumn{3}{c|}{ }   &  \multicolumn{3}{c|}{ }      \\
 & $c$ [fm$^3$] & $p$ & $\chi^2 $&  $c$ [fm$^3$] & $p$ & $\chi^2 $ \\
&&&&&&\\
\hline
\hline
&&&&&& \\
$p A$   & 5.9212 $\times 10^{0}$ & -3.7885$\times 10^{-1}$ &6.5601  &
          2.1032 $\times 10^{2}$ & -8.2425$\times 10^{-1}$ &5.7948 \\    
$\pi A$&  2.6175 $\times 10^{1}$ & -5.9388$\times 10^{-1}$ &3.8535  &
          1.6390 $\times 10^{2}$ & -8.5073$\times 10^{-1}$ &4.2729 \\
&&&&&& \\
\hline
\end{tabular}
\end{center}
\caption{Numerical results for the fit of the active volume of the gluon cloud
displayed in Fig.~\ref{volume}. For the fit a functional form 
$V = c \left(s/{\rm GeV}^2\right) ^p$ is assumed.}
\label{volumefit}
\end{table}
It is now the place to make a few statements as to the physical
meaning of  $V$ and its geometry. $V$ is the size of the gluon cloud
at the moment the interaction with the charmonium pair takes place. It should
decrease with $s$ as the faster the collision happens the less time has the cloud
to form and to expand. 
%
%
\begin{figure}[tb]
\centerline{\psfig{figure=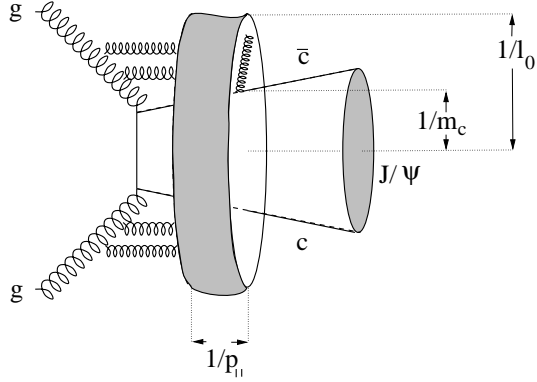,width=7cm}} 
\caption{Geometry of the cloud of hard rescattering comovers.}
\label{cloud}
\end{figure}
%
%
Fig.~\ref{cloud} shows the geometry of the cloud. The majority of all $J/\psi$ 
are produced at small $p_\perp$, so we can think us a situation where the 
$c\bar c$ pair moves essentially in the beam axis.  In transversal 
direction the cloud  should have a radius roughly comparable to 
$1/l_0 = 1/|\lvec| \sim 1/T$, its longitudinal length however depends
on a momentum $p_\parallel = \pi/(V T^2)$. 
For simplicity we did not take into account this geometry at the 'thermal' 
integration $d^3\lvec$ in Eq.~(\ref{thermal}) etc.
$p_\parallel$  will grow with $s$. 
The ratio $x_{\rm cloud}= p_\parallel/\sqrt{s}$ which is displayed in 
Fig.~\ref{xcloud}, shows what
fraction of the energy of the system is transferred to the cloud. 
Whereas the CTEQ5L gluon distribution leads to a rising fraction (which
is rather unphysical), the
GRV gluon distribution predicts more or less a constant fraction 
$x_{\rm cloud} \approx 0.5\%$ for $pA$ collisions and  $x_{\rm cloud} \approx 1\%$
for $\pi A$ collisions.
\newline\newline
%
%
%
%
We are now in a position to predict what cross section we will get with
our fit at higher energies and especially with RHIC energies with
$\sqrt{s} = 200\;{\rm GeV}$. 
%
%
\begin{figure}[tb]
\centerline{\psfig{figure=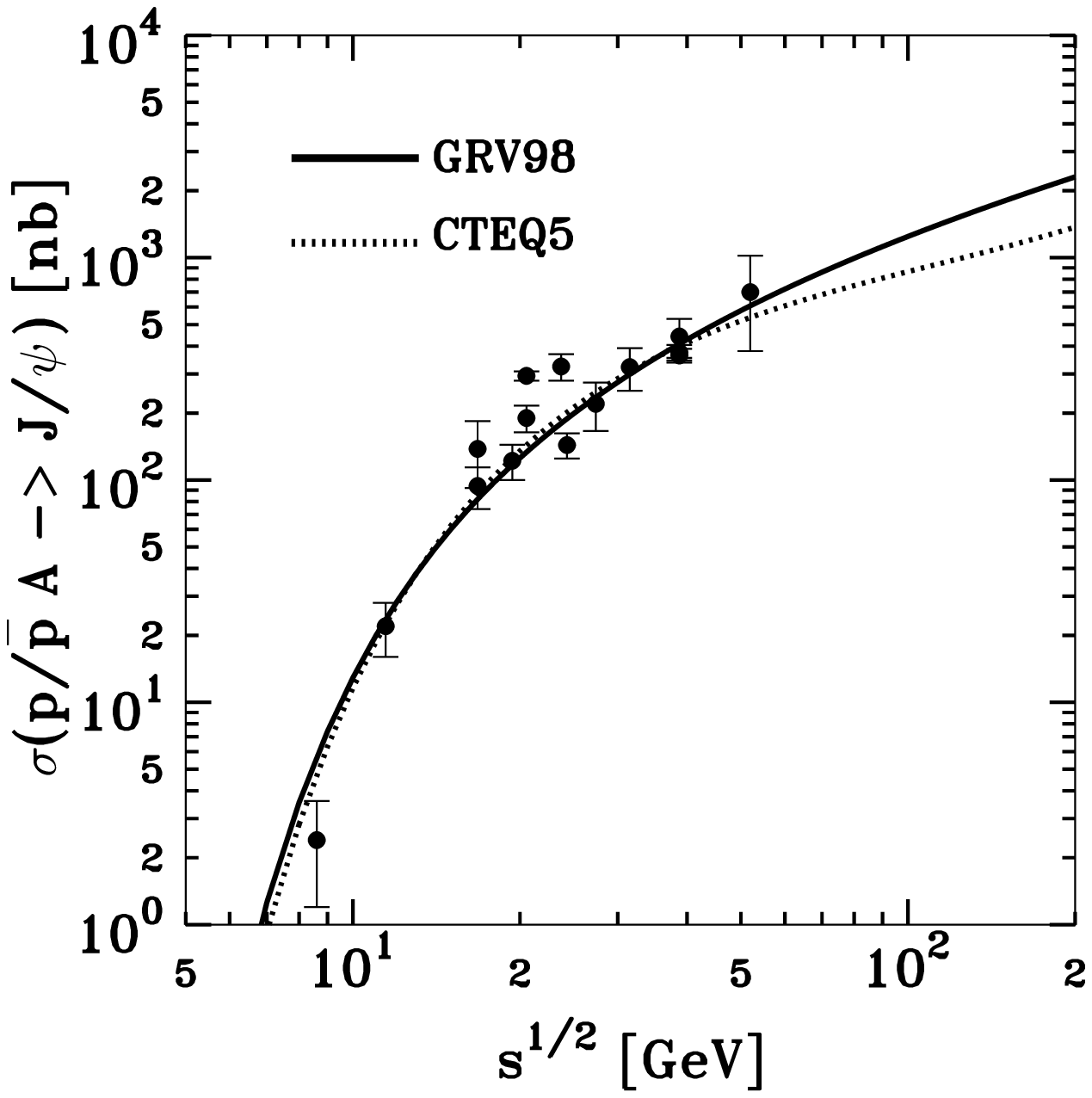,width=6.5cm}
            \psfig{figure=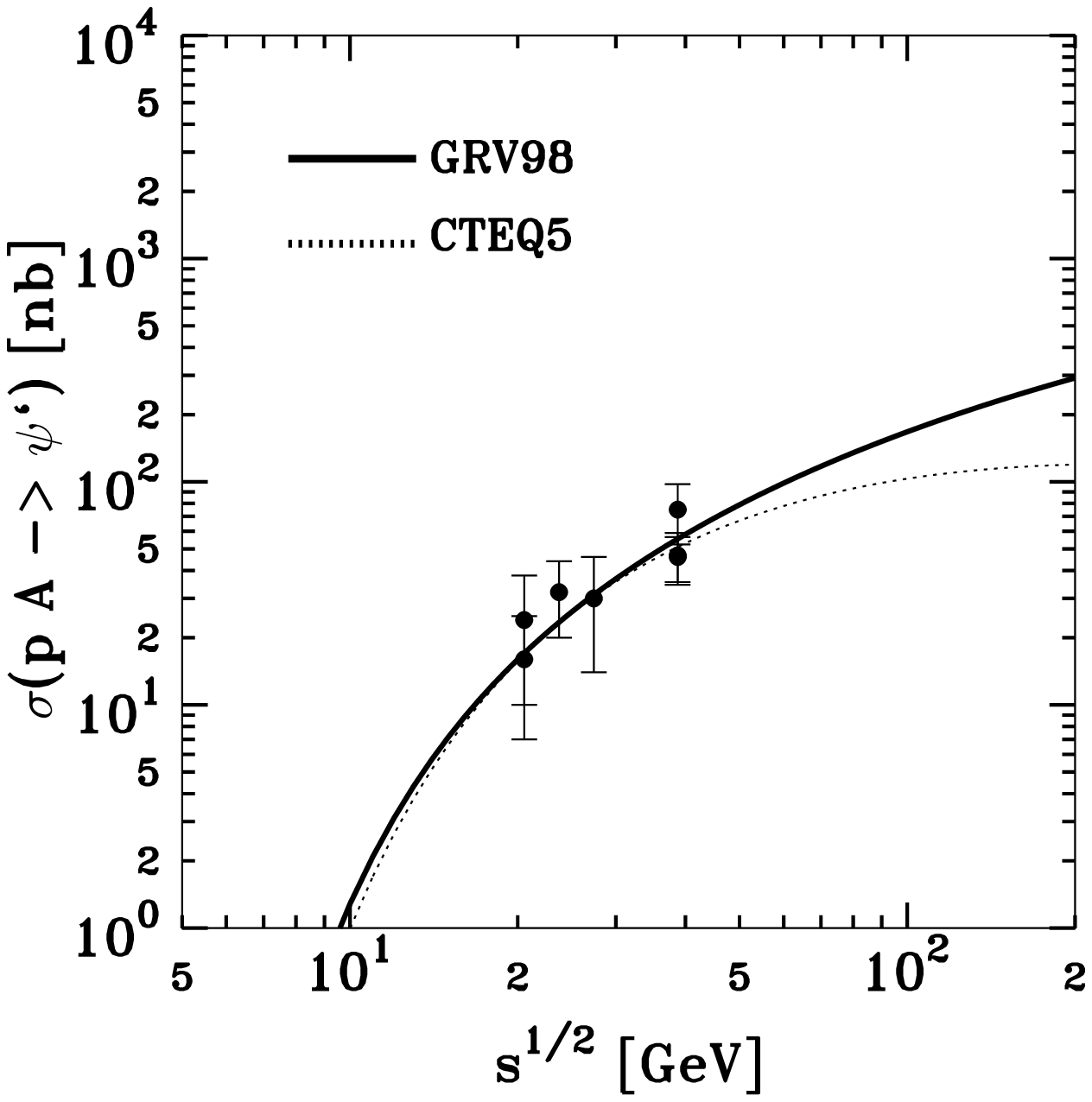,width=6.5cm}} 
\centerline{\psfig{figure=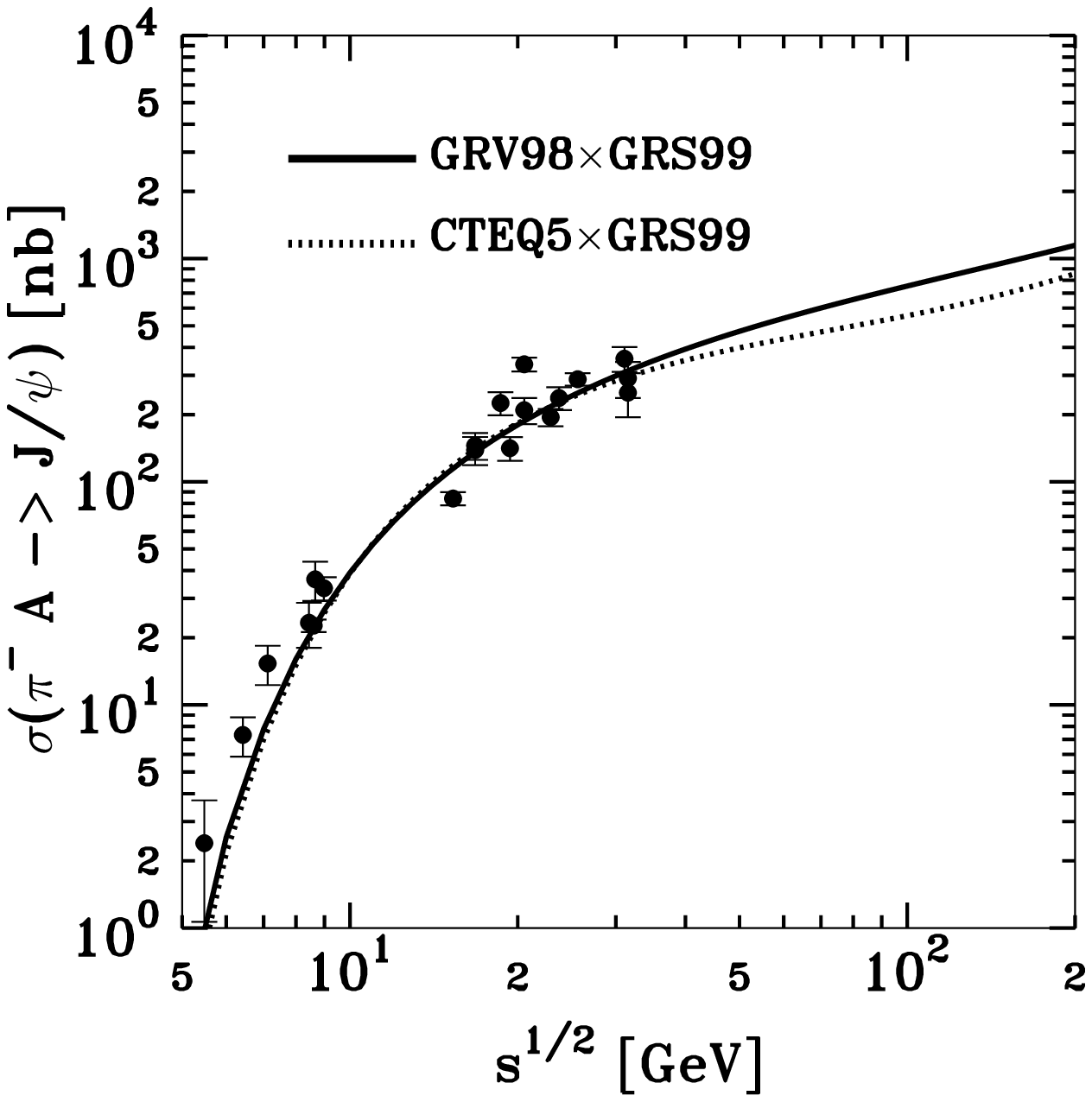,width=6.5cm}
            \psfig{figure=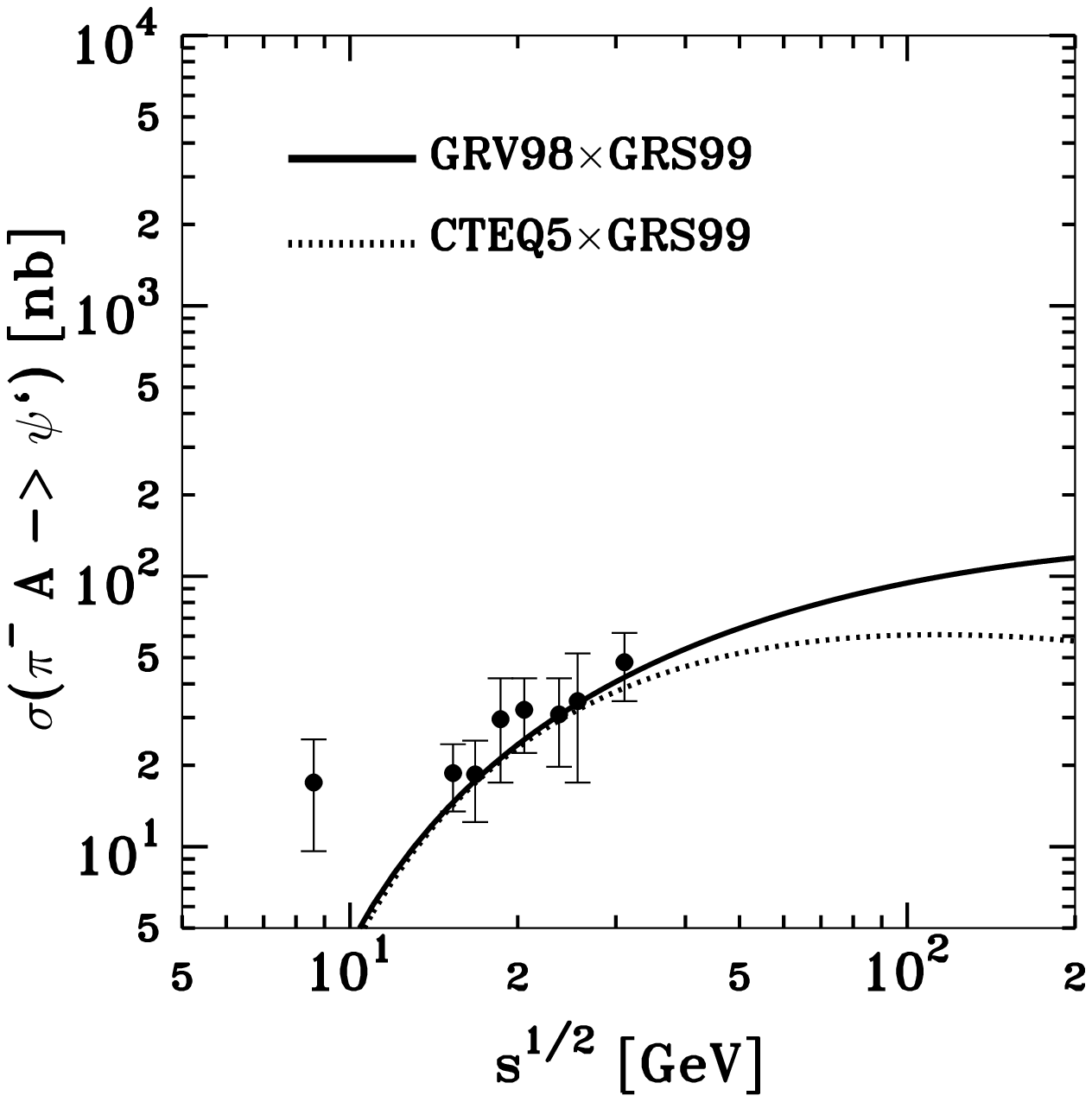,width=6.5cm}} 
\caption{Total cross section for S-wave charmonium production $\sigma$  
in nb versus $\sqrt{s}$: On the left the cross sections for $J/\psi$ production 
are displayed and on the right the ones for $\psi'$ production. The upper two 
figures show the results for $pA$ and the lower two figures for $\pi A$ 
collisions. All nuclear effects have been rescaled so that in principle all 
the cross sections should display the result for $pp$ and $\pi p$ collisions, 
respectively. The solid and the dashed line show the curves obtained from 
the combined ($J/\psi$ and $\psi'$ data) fit for the temperature
depending on which parton distribution has been used. 
}
\label{datasample}
\end{figure}
%
%
Fig.~\ref{datasample} shows the fit we made in terms of the total inclusive
$J/\psi$ and $\psi'$ cross section. 
The data have been rescaled
to use the full range in $x_F\in [-1,1]$. For the $\pi A$ collisions we have assumed
an $x_F$ distribution of the form $d\sigma/dx_F \sim (1-|0.18-x_F|)^c$, with
$c= 2.5$ for $J/\psi$ and  $c=3.9$ for $\psi'$ \cite{Schuler:1994hy}. 
In case of the $pA$ collisions a symmetric $x_F$ distribution is assumed.
It turns out that for the latter one the
CTEQ5L distribution predicts an unphysical decreasing cross section at large
$\sqrt{s}$ therefore we will not consider this distribution in the following
any longer, whereas GRV shows in all cases a reasonable relaxing rising 
behavior. We can now have a more closer look at the details of the various
processes contributing using the GRV set only. The first important 
quantity of interest in the $\chi_1/\chi_2-{\rm ratio}$. It is defined
by:
\begin{equation}
\chi_1/\chi_2-{\rm ratio} = 
\frac{\sigma(\chi_{c1}){\rm Br}(\chi_{c1}\to J/\psi)}
     {\sigma(\chi_{c2}){\rm Br}(\chi_{c2}\to J/\psi)}\;.
\end{equation}
A comparison of the data and our results is shown in Tab.~\ref{chi12ratio}.
\begin{table}
\begin{center}
\begin{tabular}{|c||c|c|c|}
\hline &&& \\
Reference &  E beam [GeV] & $(\chi_1/\chi_2)_{\rm exp.}$ &  $(\chi_1/\chi_2)_{\rm theor.}$ \\ 
&&& \\
\hline \hline 
&&& \\
E705   \cite{Antoniazzi:1993iv} & 185 & 1.4 $\pm$ 0.4  & 1.0456  \\
E506   \cite{Koreshev:1996wd}   & 515 & 1.2 $\pm$ 0.4  & 0.9194  \\
&&&\\
\hline
\end{tabular}
\end{center}
\caption{ Comparison of the measured $\chi_1/\chi_2-{\rm ratio}$ in $\pi A$ reaction with the result of the theory described here. For the gluon distribution
function the GRV98 $\times$ GRS99 set is used.}
\label{chi12ratio}
\end{table}
It turns out that the values for the $\chi_1/\chi_2-{\rm ratio}$ are a bit smaller,
but well inside the error bars of the
 measured  $\pi A$ reactions. They would be reproduced even better, if we neglected
the CSM contributions altogether. In this case we get $(\chi_1/\chi_2)_{\rm theor.} = 1.2$
independent of the beam energy, if we set all masses equal to $2m_c$. 
So, in principle the theory is capable to describe the large $\chi_1/\chi_2-{\rm ratio}$
found experimentally in contrast to the standard COM and CSM theory.
\newline \newline
%
%
\begin{figure}[tb]
\centerline{\psfig{figure=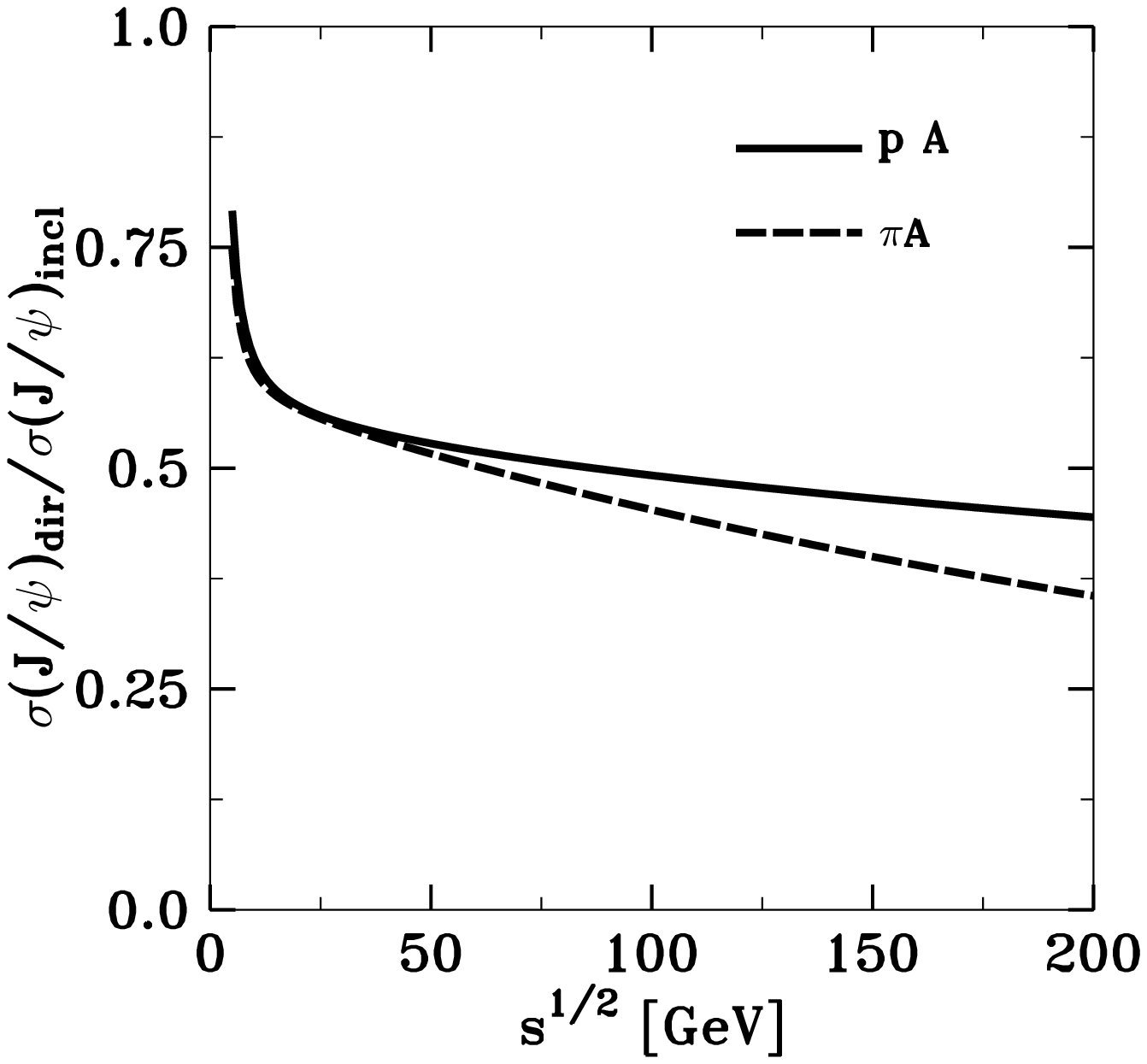,width=6cm}
            \psfig{figure=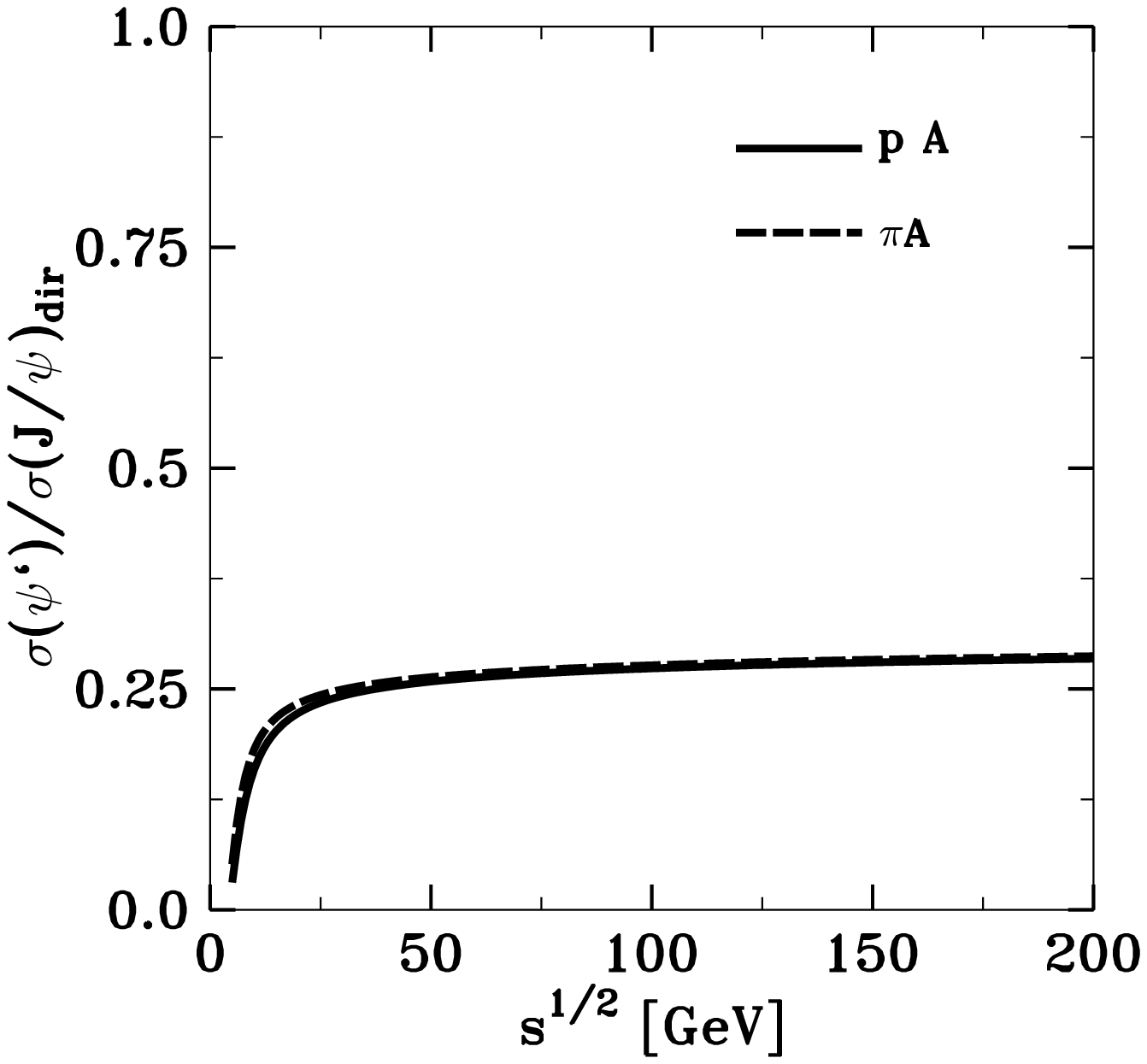,width=6cm}
            \psfig{figure=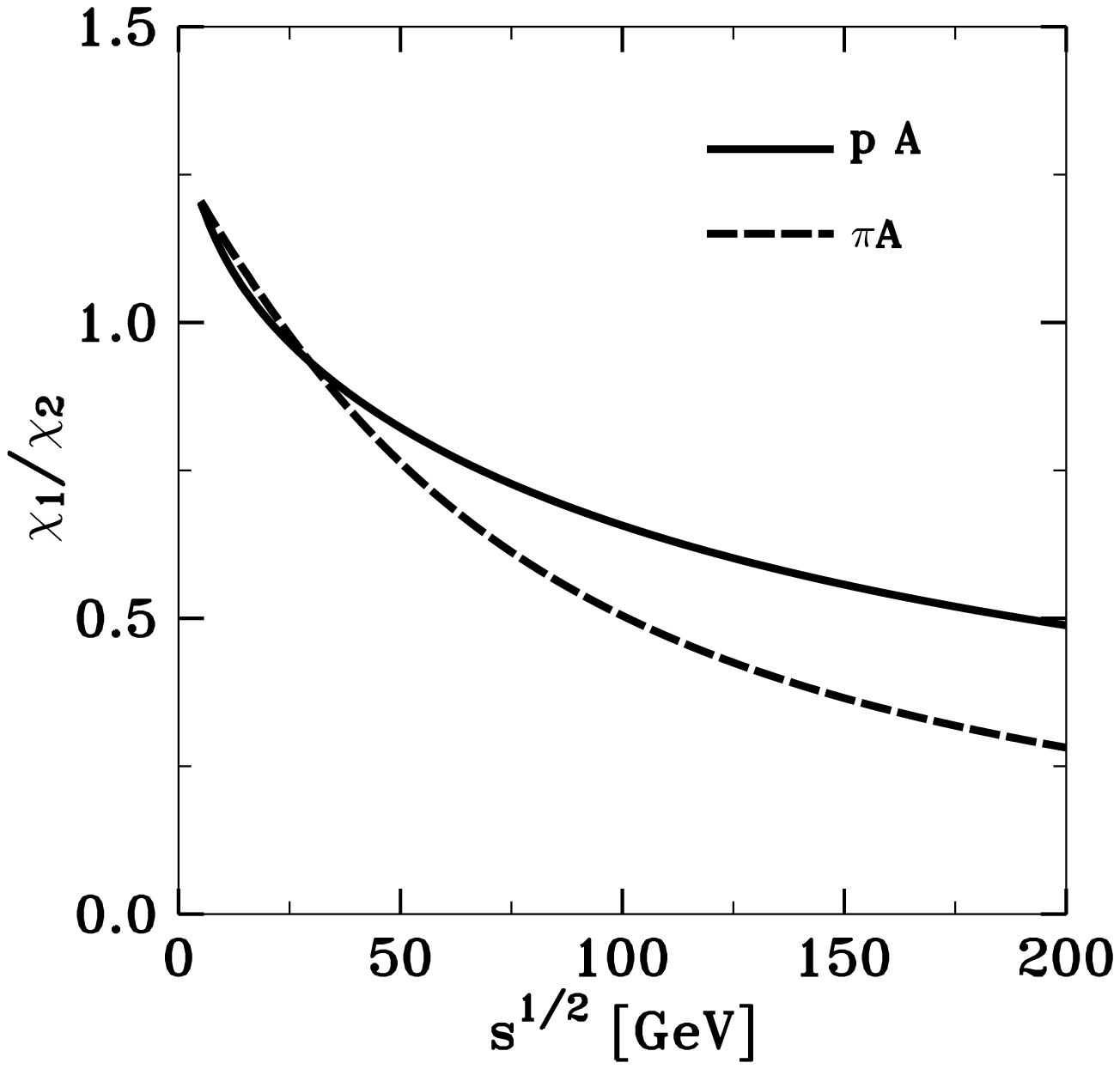,width=6cm}} 
\caption{Contribution of various subprocesses to inclusive $J/\psi$ production 
in $pA$ and $\pi A$ collisions. On the right the ratio of the directly produced
$J/\psi$ to all  $J/\psi$ measured is displayed. In the middle the ratio between 
the $\psi'$ cross section and the cross section of the directly produced $J/\psi$
is shown. The shape of this ratio changes with $s$ only due to the evolution
of the gluon parton density. On the right the ratio $\chi_{c1}/\chi_{c2}$ as discussed
in the text is displayed. The underlying gluon distribution for all figures has
been taken from GRV and GRS respectively.}
\label{details}
\end{figure}
%
%
Fig.~\ref{details} shows some details as to the various subprocesses contributing
to the inclusive $J/\psi$ production in $p A$ and $\pi A$ scattering. It is shown
that the fraction of directly produced $J/\psi$ versus the whole inclusive
cross section decreases with increasing energy until it falls down to about 40\%
at RHIC energies. The $s$ dependence of the $\sigma(\psi')/\sigma(J/\psi)_{\rm dir}$
is governed by the scale dependence of the gluon parton distribution (here GRV) involved.
It goes towards a constant for large $s$ which is about 1/4. The $\chi_{c1}/\chi_{c2}$
ratio is rapidly falling with $s$. In all cases the $p A$ and $\pi A$ curves behave
similar.
\section{Double spin asymmetries}
\label{secall}
The discussion on the partonic cross section has shown two important consequences:
\begin{itemize}
\item As long as we can set the displacement parameter $b$ to zero,
which in accordance to the non-polarization of the final $J/\psi$, there
is no correlation between the proton spin and the final $J/\psi$ spin
orientation. All single spin asymmetries are then zero as to
the order of accuracy of the approximations applied here. 
\item The double spin asymmetries for the directly produced $J/\psi$ and $\psi'$
depend up to a factor only on the ratio of the polarized and unpolarized 
gluon distributions. In case of the inclusive  $J/\psi$ cross section the 
different mass-scales make the situation more complicated.
\end{itemize}
These findings mean a big simplification for the polarized physics because they state that the extraction of the polarized gluon density from the 
double spin asymmetry  will not be complicated by 
initial and final state spin correlations. With the model set up in the previous chapter
we are able to compute the error bars for the double spin asymmetries for $J/\psi$, $\psi'$
and $\chi_{cJ}$ production. In the following we collect the expressions for the
double longitudinal spin cross section $\Delta \sigma_{LL}$:
\begin{eqnarray}
\Delta \sigma_{LL}  &=& -\int dx_1 dx_2 \sum_{S_z}\Sigma_{LL}(S_z) 
\quad ({\rm general\; formula})
\nonumber \\
\Delta \sigma_{LL}(J/\psi)_{\rm dir} &=&  3\frac{V \zeta(4)}{2 \pi^2 \beta^4}
\int dx_1 \int dx_2 {\cal F}_S( M_{J/\psi}^2)
\Delta G(x_1, M_{J/\psi}^2) \Delta G(x_2, M_{J/\psi}^2)
\nonumber \\
\Delta \sigma_{LL}(\psi') &=&  3\frac{V \zeta(4)}{2 \pi^2 \beta^4}
\int dx_1 \int dx_2 {\cal F}_S( M_{\psi' {\rm eff}}^2)
\Delta G(x_1, M_{\psi' {\rm eff}}^2) \Delta G(x_2, M_{\psi' {\rm eff}}^2)
\nonumber \\
\Delta \sigma_{LL}(\chi_{cJ}') &=& (2J+1) \frac{V \zeta(4)}{2 \pi^2 \beta^4}
\int dx_1 \int dx_2 {\cal F}_P( M_{\chi_{cJ} {\rm eff}}^2)
\Delta G(x_1, M_{\chi_{cJ} {\rm eff}}^2) \Delta G(x_2, M_{\chi_{cJ} {\rm eff}}^2)
\nonumber \\
&& + \Delta  \sigma_{LL}^{\rm (CSM)}(\chi_{cJ})
\nonumber \\
\Delta \sigma_{LL}(J/\psi)_{\rm incl} &=&
\Delta \sigma_{LL}(J/\psi)_{\rm dir}
+        {\rm Br}( \psi' \to J/\psi) \Delta \sigma_{LL}(\psi')
+\sum_J  {\rm Br}( \chi_{cJ} \to J/\psi) \Delta \sigma_{LL}(\chi_{cJ})\;.
\nonumber \\
\end{eqnarray}
The - sign in the general formula takes into account that the standard
convention for the numerator of the asymmetry is always anti-parallel
spin alignment minus parallel spin alignment.
For the color singlet (CSM) contributions we find:
\begin{eqnarray}
\Delta  \sigma_{LL}^{\rm (CSM)}(\chi_{c0}) &=& \frac{12 \pi^2 \alpha_s^2 |R_1'|^2}
{ M_{\chi_{c0} }^7} \int dx_1 dx_2 
\Delta G(x_1, M_{\chi_{c0} }^2) \Delta G(x_2, M_{\chi_{c0} }^2)
\delta\left(1- \frac{M_{\chi_c0}^2}{S x_1 x_2}\right)
\nonumber \\
\Delta  \sigma_{LL}^{\rm (CSM)}(\chi_{c1}) &=& 0
\nonumber \\
\Delta  \sigma_{LL}^{\rm (CSM)}(\chi_{c2}) &=&
 -\frac{16 \pi^2 \alpha_s^2 |R_1'|^2}
{ M_{\chi_{c2} }^7} \int dx_1 dx_2 
\Delta G(x_1, M_{\chi_{c2} }^2) \Delta G(x_2, M_{\chi_{c2} }^2)
\delta\left(1- \frac{M_{\chi_c2}^2}{S x_1 x_2}\right)\;.
\nonumber \\
\end{eqnarray}
The corresponding unpolarized cross sections can be straightforwardly obtained
by replacing the polarized by the unpolarized gluon distributions and to remove
all minus signs. Then the double spin asymmetry 
$A_{LL}$ and its statistical error $\delta A_{LL}$ 
are simply given by:
\begin{equation}
A_{LL} = \delta \sigma_{LL}/\sigma\;; \qquad \delta A_{LL} = 2
\sqrt{\frac{\sigma_+ \sigma_-}{{\cal L}(\sigma_++\sigma_-)^3}}
\;, \quad \sigma_\pm = \sigma\pm \Delta \sigma_{LL}\;.
\end{equation}
%
%
\begin{figure}[tb]
\centerline{\psfig{figure=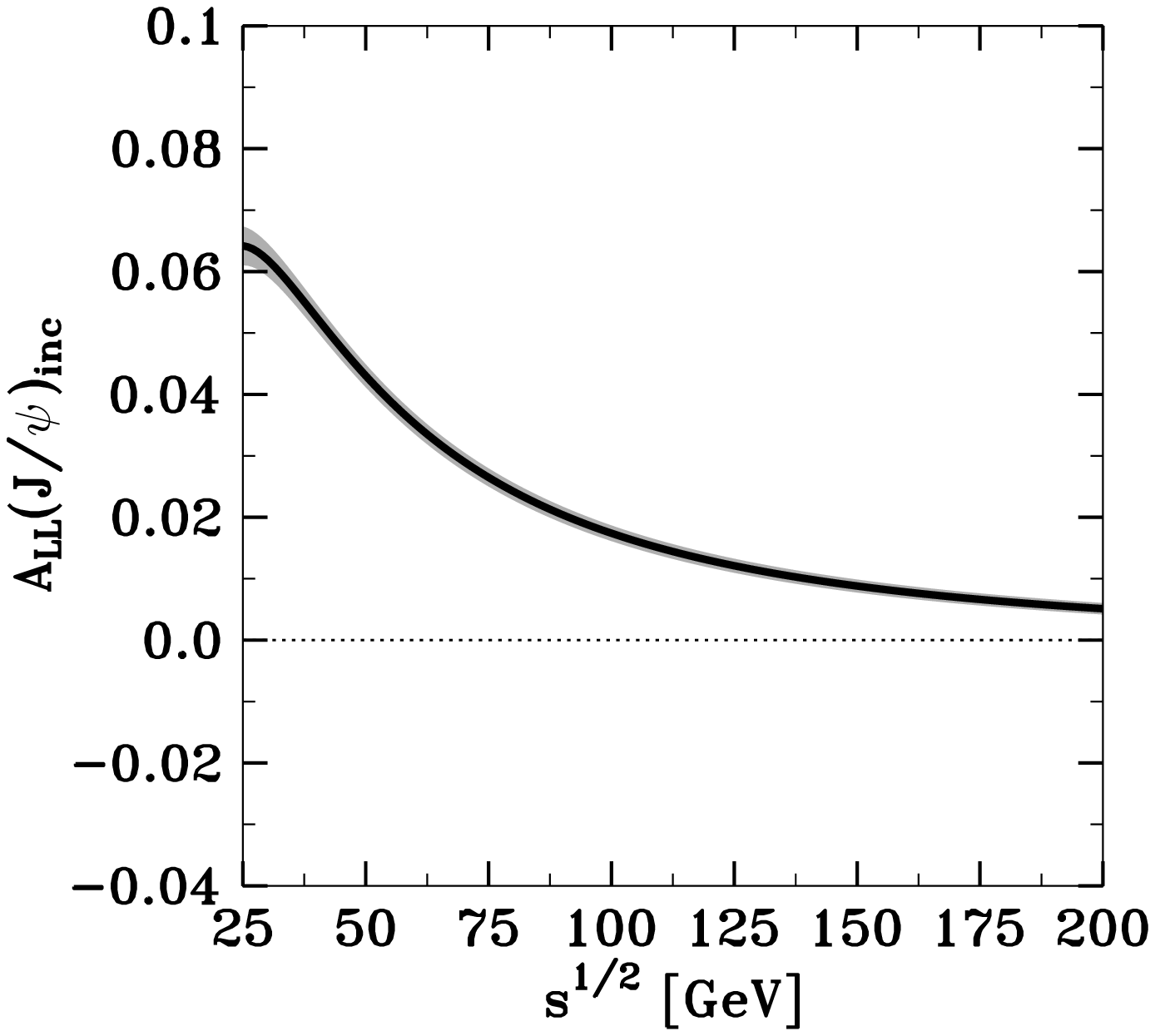,width=7cm} 
            \psfig{figure=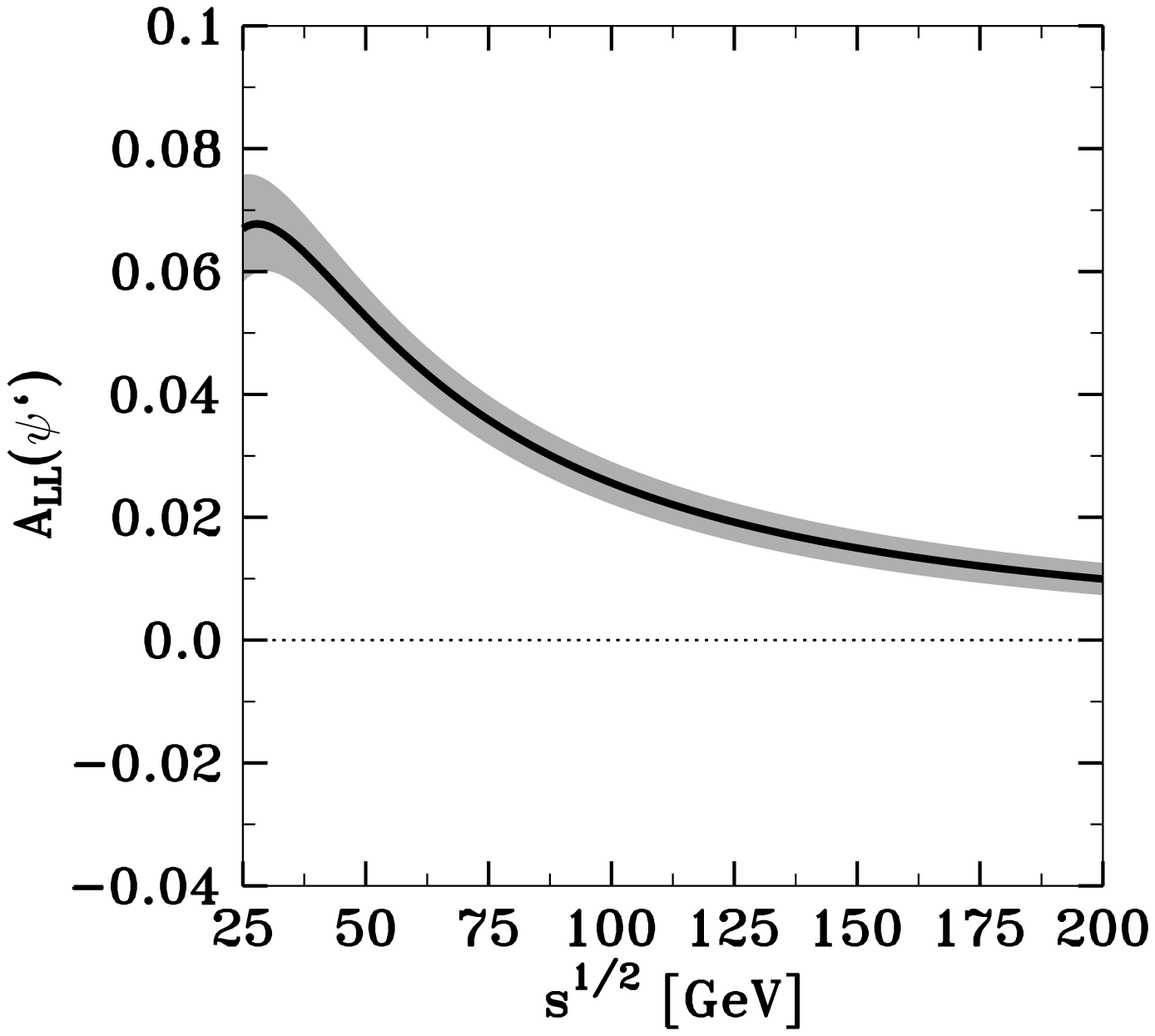,width=7cm}} 
\caption{Double spin asymmetry $A_{LL}$ for inclusive $J/\psi$ (left) 
and $\psi'$ (right)  hadroproduction. For the plot we use the unpolarized
parton distribution set GRV and the polarized GSA. For the grey error band 
the assumed luminosity
is $0.25\; {\rm pb}^{-1}$ using a beam polarization of 100\%.}
\label{allswave}
\end{figure}
\begin{figure}[tb]
\centerline{\psfig{figure=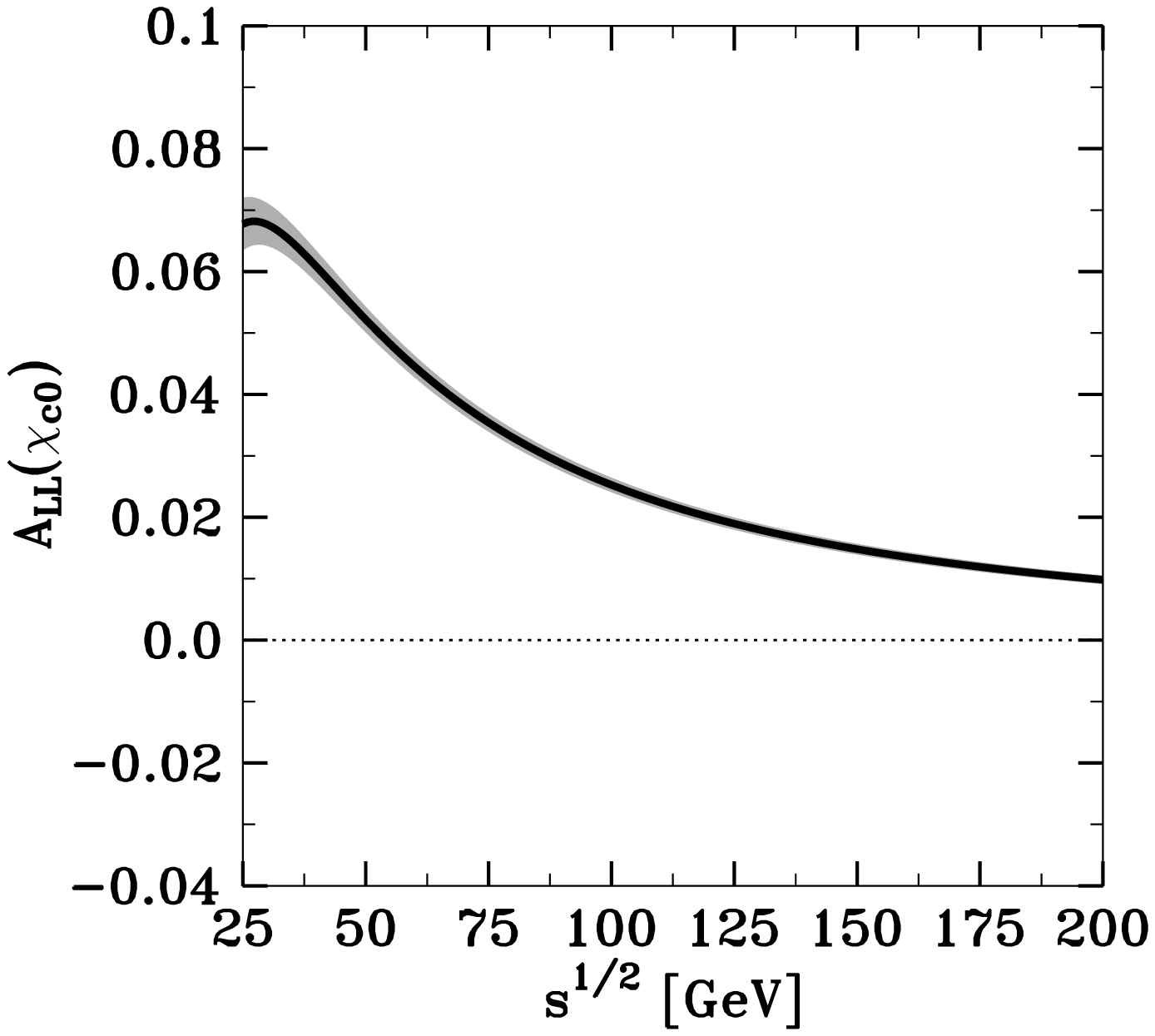,width=6cm} 
            \psfig{figure=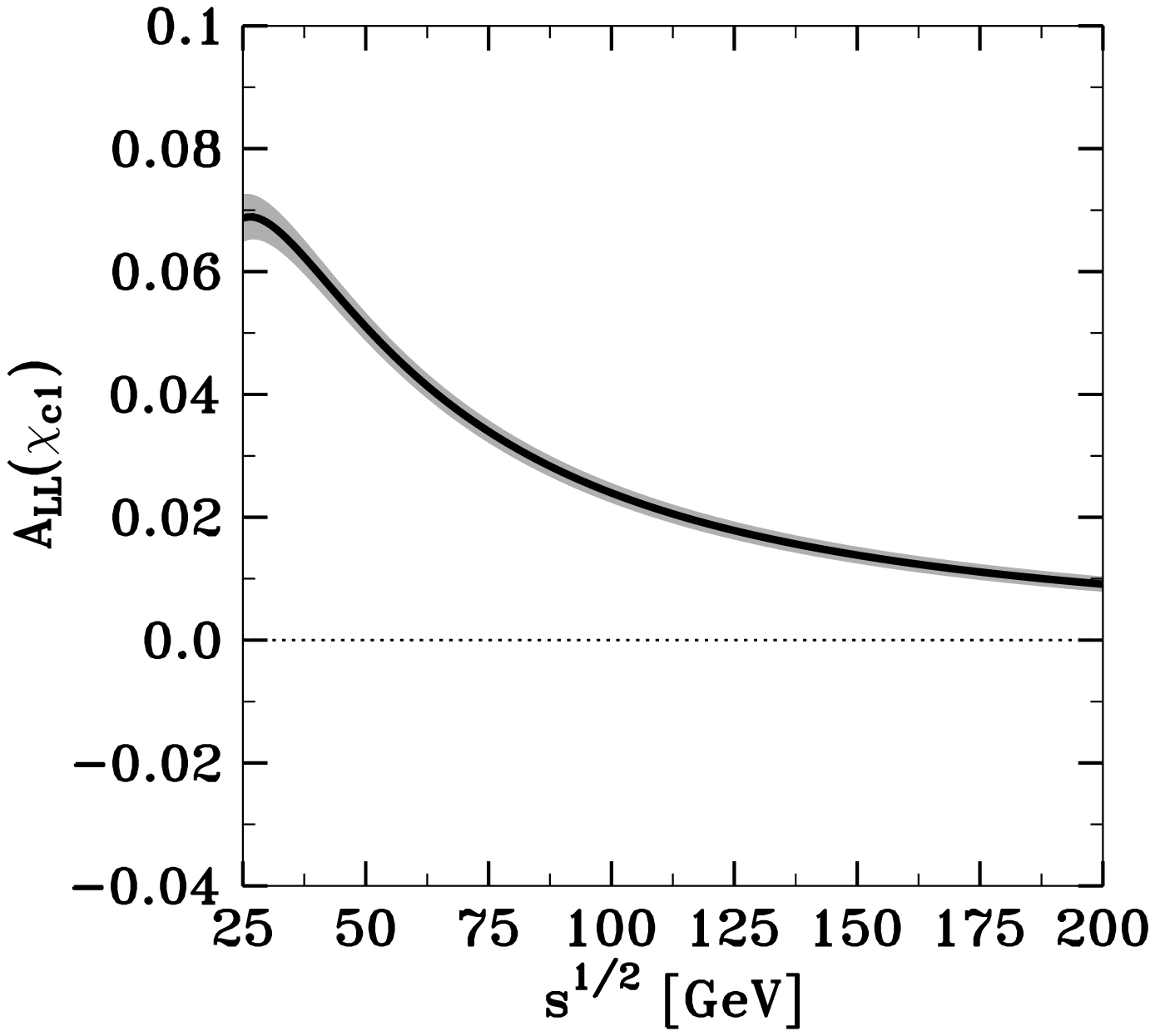,width=6cm} 
            \psfig{figure=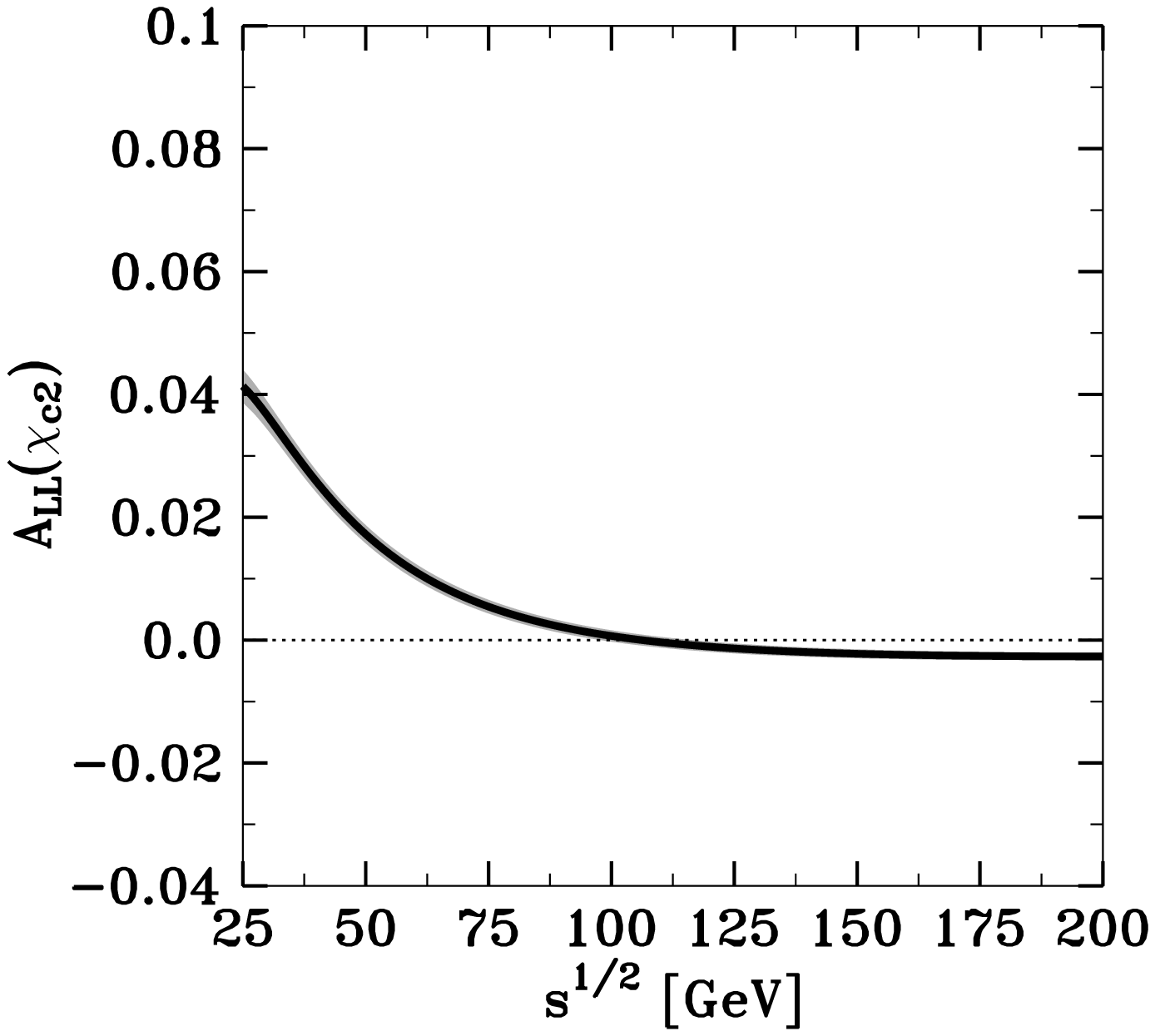,width=6cm}} 
\caption{Double spin asymmetry $A_{LL}$ for inclusive $\chi_{cJ}, J=0,1,2$  
hadroproduction. For the plot we use the unpolarized
parton distribution set GRV and the polarized GSA. For the grey 
error band the assumed luminosity
is $0.25\; {\rm pb}^{-1}$ using a beam polarization of 100\%.}
\label{allpwave}
\end{figure}
%
Fig.~\ref{allswave} shows the double spin asymmetry for S-wave charmonium production,
i.e inclusive $J/\psi$ and $\psi'$ production. For the polarized gluon distribution
amplitude we use the leading order set gluon A \cite{Gehrmann:1996ag}, which we
will abbreviate in the following by GSA.
For the error band we have assumed
a luminosity of ${\cal L}=0.25\;{\rm pb}^{-1}$. 
It is seen that  for inclusive $J/\psi$ and $\psi'$ production the asymmetry
at RHIC energies is sizable. It is larger for $\psi'$ production, but here also
the error bars are larger.
For P-wave charmonium production (see Fig.~\ref{allpwave})
the asymmetry for the $\chi_{c2}$ production is partially negative due to the  CSM
contribution. In case the CSM contribution is smaller, i.e. that
a smaller value for $|R_1'|$ is more realistic, the asymmetry will go
to more positive values, so here we see a very fine test of the interplay
between CSM and comovers. In all cases we get a substantial asymmetry 
for RHIC energies with small error bars. In general it is noticed that
the asymmetry decreases with increasing beam energy $\sqrt{s}$.
This means that in addition to the RHIC spin program
an polarized experiment like HERA-$\vec N$, which is supposed to run
at $\sqrt{s} = 40 \;{\rm GeV}$ will also contribute very valuable information
for the mechanism how charmonium production happens in hadroproduction. 
One should notice that the asymmetries for inclusive $J/\psi$
and the $\chi_{c1}$ production do not depend on the temperature while
the other asymmetries do, so to measure the asymmetries will give 
essential new information upon the validity of the theory.
\newline \newline
Finally, it has to be stated that the whole
calculation still is based on the velocity expansion $p/m$ 
within the NRQCD formalism, which is 
truncated already at the first order. The corrections to this may be
quite considerable. In this direction we find  also the big
expansion parameter $\rho$ already determined in \cite{Hoyer:1999ha} and
which has been confirmed here. Unfortunately, the inclusion of those
velocity corrections will destroy the simple
relations derived in \cite{Hoyer:1999ha,Marchal:2000wd}. It would
be quite advisable for future studies to develop a formalism,
that could test the results of the hard comover rescattering from
a different stand point which is not based on the approximations
of the NRQCD. 
\section{Summary}
In this work we have tried to describe the measured unpolarized
cross section for charmonium production through the framework
of hard comover rescattering and made predictions for the asymmetries
in polarized $pp$ scattering. The generic advantage of the hard 
comover rescattering mechanism is that it can explain the non-polarization
and the comparatively large value for the $\chi_1/\chi_2$-fraction
observed in experiment in a simple and natural way.
\newline
\newline
In order to get quantitative results we
have expressed the comovers as a thermal cloud of gluons. The measured
data suggest that  about 0.5-1\% of the total 
energy in the collision is invested in the formation of the gluon 
cloud. The fact,
that the final state $J/\psi$ is unpolarized leads us to the
conclusion that the displacement parameter $b$ should be small and
consistent to zero. If this is true then it means that there is no
correlation in polarized experiments between the initial polarization
of the protons and the final polarization of the $J/\psi$ and
furthermore the single-spin asymmetry $\Sigma_{0L}$ should be zero,
a notion which should be tested by experiment. 
The asymmetries $A_{LL}(\psi')$ and $A_{LL}(\chi_{c1})$
depend only on the ratio of the polarized and unpolarized gluon distribution
amplitudes, while $A_{LL}(J/\psi)_{\rm incl}$,   $A_{LL}(\chi_{c0})$
and $A_{LL}(\chi_{c2})$ are in principle sensitive to the parameters
describing the gluon cloud. The hard comover rescattering picture
provides an  understanding of the formation of onium states
in hadroproduction which may give the answers to some problems
left unsolved in the standard COM and CSM mechanism. The RHIC-spin
experiment
and a possible HERA-$\vec N$ will provide very crucial new information
upon the validity and the consequences of the theory presented here.
\newpage
\par
\noindent
{\Large\bf Acknowledgments}
\newline
It is a pleasure to thank Paul Hoyer,
Nils Marchal  and St\'ephane Peign\'e for fruitful
discussion and many explanations concerning their work. I also
want to thank Marc Bertini and Torbjorn Sj\"ostrand 
for helpful suggestions and ideas.
\vskip0.5cm\par\noindent

\newpage
\begin{appendix}
\section{Derivation of the $\chi_{0}$ and $\chi_{2}$ ${\cal O}(\as^2)$
\label{app1}
CSM cross sections}
In this appendix we reproduce the Born cross section for  
$\chi_{0}$ and $\chi_{2}$ production (${\cal O}(\as^2)$). This gives
a cross check for the formulas derived in \cite{Hoyer:1999ha} and
also a cross check for the phase-space  and flux 
factors we used in the text to obtain the cross section formulas
from the amplitudes.
\newline \newline
The momenta for the gluon fusion amplitude are:
\begin{eqnarray}
g_1 = (m,0,0,m), && \quad g_2 = (m,0,0,-m) \nonumber \\
p_1 = (m,{\bf p}) , && \quad p_2 =  (m,-{\bf p})
\end{eqnarray}
\begin{eqnarray}
\Phi &=& -ig^2 \bar v(p_2,\bar \lambda) 
\left[ \epsla_2 T^b \frac{\psla_1 - \gsla_1 + m_Q}{(p_1-g_1)^2 - m_Q^2}
       \epsla_1 T^a
+
 \epsla_1 T^a \frac{\psla_1 - \gsla_2 + m_Q}{(p_1-g_2)^2 - m_Q^2}
       \epsla_2 T^b \right] u(p_1,\lambda)
\nonumber \\
 && + \frac{-g^2}{(2m)^2} f_{abc} \;\epsvec_1 \cdot \epsvec_2 \;  
 \bar v(p_2,\bar \lambda) ( \gsla_1 -\gsla_2)  u(p_1,\lambda)
\nonumber \\
\nonumber \\
& \sim &
\frac{ig^2}{2m} \bar v(p_2,\bar \lambda)
\left[ \epsla_2 T^b (\psla_1 - \gsla_1 + m_Q)  \epsla_1 T^a 
\left(\frac{1}{m} +\frac{p_z}{m^2} \right)\right.
\nonumber \\ && \qquad \qquad  \qquad \left.
+
 \epsla_1 T^a (\psla_1 - \gsla_2 + m_Q)  \epsla_2 T^b 
\left(\frac{1}{m} -\frac{p_z}{m^2} \right) \right]  u(p_1,\lambda)
\nonumber \\
 && + \frac{g^2}{2m} f_{abc}\; \epsvec_1 \cdot \epsvec_2 \;  
 \bar v(p_2,\bar \lambda) \gamma^3  u(p_1,\lambda)\;.
\end{eqnarray}
Using now the equation of motion one can write:
\begin{eqnarray}
 \epsla_2 T^b (\psla_1 - \gsla_1 + m_Q)  \epsla_1 T^a 
&=& -T^b T^a \left[
    \epsla_1 \epsvec_2\cdot \pvec 
   + \epsla_2 \epsvec_1\cdot \pvec 
+m\left( \epsvec_1 \cdot \epsvec_2 \gamma_3 
     +i [\epsvec_1\times \epsvec_2]^z \gamma_0 \gamma_5 \right) 
\right]
\nonumber \\
 \epsla_1 T^a (\psla_1 - \gsla_2 + m_Q)  \epsla_2 T^b 
&=& -T^a T^b \left[
    \epsla_1 \epsvec_2\cdot \pvec 
   + \epsla_2 \epsvec_1\cdot \pvec 
-m\left( \epsvec_1 \cdot \epsvec_2 \gamma_3 
     -i [\epsvec_1\times \epsvec_2]^z \gamma_0 \gamma_5 \right) 
\right]\;.
\nonumber \\
\end{eqnarray}
The spinor combinations can be expressed as follows:
\begin{eqnarray}
 \bar v(p_2,\bar \lambda) \gamma^3  u(p_1,\lambda) &=& 2m 
\delta^{-\bar \lambda}_\lambda 2\lambda
\nonumber \\
\bar v(p_2,\bar \lambda) \gamma^0\gamma_5  u(p_1,\lambda) &=& 2m 
\delta^{-\bar \lambda}_\lambda 
\nonumber\\
\bar v(p_2,\bar \lambda) \epsla  u(p_1,\lambda) &=& 2m 
\delta^{\bar \lambda}_\lambda (-2\lambda)( \evec^*(2\lambda) \cdot \epsvec) 
\sqrt{2}\;.
\end{eqnarray}
Then, using $[T^a,T^b]= if_{abc}T^c$ and 
      $\{T^a,T^b\} = d_{abc}T^c$,
we arrive at the following expression in the first order $p/m$:
\begin{eqnarray}
\Phi &=& -i g^2 
\left\{ 
i \delta^{-\bar \lambda}_{\lambda} [\epsvec_1 \times \epsvec_2]^z
\left[
\left( \frac{1}{N_c} \delta_{ab}\delta_{ij}
+ d_{abc} T^c_{ij} \right) 
-\frac{ip_z}{m} f_{abc} T^c_{ij} \right] \right.
\nonumber 
\\ && \qquad \left.
-2\lambda \delta^{\bar \lambda}_{\lambda} \sqrt{2}
\left(
(\evec^*(2\lambda)\cdot \epsvec_1) (\epsvec_2 \cdot \pvec)
+( \evec^*(2\lambda)\cdot \epsvec_2) (\epsvec_1 \cdot \pvec) \right)
\frac{1}{m}
\left( \frac{1}{N_c} \delta_{ab} \delta_{ij}
+ d_{abc} T^c_{ij} \right) 
 \right\}
\nonumber 
\\ && \qquad
+ 2\lambda   \delta^{-\bar \lambda}_{\lambda}\epsvec_1 \cdot \epsvec_2
 \frac{p_z}{m}
\left( \frac{1}{N_c} \delta_{ab}\delta_{ij}
+ d_{abc} T^c_{ij} \right)\;.  
\end{eqnarray}
Hereby we reproduce up to convention dependent phase factors
Eq.~(4) in \cite{Hoyer:1999ha}.
For the wave function one uses the following expression:
\beq
\psi_{\lambda\bar \lambda}^{L_z S_z}(\qvec) = \psi_{LL_z} \frac{1}{\sqrt{2}}
\evec(S_z) \chi_{-\bar\lambda} {\bf \sigma} {\chi}_{\lambda}\;.
\eeq
Now the amplitude for $\chi_{cJ}, J=0,1,2$ production is given by:
\beq
{\cal M}^{(a)}(^3P_J,J_z) = \sum_{\lambda \bar \lambda \atop L_z S_z}
\left( \begin{array}{cc} 1 & 1 \\
                         L_z & S_z 
       \end{array} \right| \left. 
\begin{array}{c} J \\ J_z \end{array} \right)
\int\frac{d \qvec}{(2\pi)^3} \phi^{[1]}_{\lambda \bar \lambda}(\qvec) 
                            \psi_{\lambda \bar \lambda}^{L_z S_z}(\qvec)\;.
\eeq
With this we find for the $\chi_{c0}$ meson:
\beq
{\cal M}^{(a)}(^3P_0,0) = -g^2 \frac{1}{N_c}\delta_{ij} \delta_{ab}
\frac{3}{\sqrt{2\pi m^3}} R_1' \delta_{\lambda_1}^{-\lambda_2}\;, 
\eeq
with $R_1'$ being the first derivative of the quarkonium
wave function at the origin, as defined by:
\beq
\int\frac{\qvec d^3 \qvec }{(2\pi)^3}  \psi_{LL_z}(\qvec)
= i \sqrt{\frac{3}{4\pi m}} R_1' \evec(L_z)\;.
\eeq
Then the partonic cross section is given by:
\beq
\sigma_{\lambda_1 \lambda_2}(^3P_0,0)_{CSM}
= \frac{2\pi}{2(2m)^4} \frac{1}{\left(N_c^2-1\right)^2}\sum_{ab} 
\frac{1}{N_c^2} \sum_{ij}
|{\cal M}^{(a)}(^3P_0,0)|^2 
= \frac{24\pi^2 \as^2 |R_1|^2}{(2m)^7}  \delta_{\lambda_1}^{-\lambda_2} 
\;.
\eeq
The amplitude for the $\chi_{c1}$ meson vanishes identically. For
the $\chi_{c2}$  meson we find then:
\beq
{\cal M}^{(a)}(^3P_2,J_z) = -g^2 \frac{2}{N_c}\delta_{ij} \delta_{ab}
\sqrt{\frac{3}{2\pi m^3}} R_1'  \delta_{\lambda_1}^{\lambda_2}
\delta_{J_z/2}^{\lambda_1} \;,
\eeq
and we obtain for the partonic cross section henceforward:
\beq
\sigma^{(a)}_{\lambda_1 \lambda_2}(^3P_2,J_z)_{ \rm CSM}
= \frac{32\pi^2 \as^2 |R_1|^2}{(2m)^7}  \delta_{\lambda_1}^{\lambda_2} 
\delta_{J_z/2}^{\lambda_1} \;.
\eeq
\section{On the alternative of a virtual gluon field $\Gamma$}
\label{virtreal}
A virtual gluon field can be parameterized by the transversality condition:
\begin{equation}
\Gamma_\mu \Gamma_{\mu'}^* = g_{\mu\mu'} - \frac{l_\mu l_{\mu'}}{l^2}\;.
\end{equation}
Taking now the incoming and outgoing charm quark to be on-shell one obtains
up to velocity corrections:
\begin{equation}
m_c^2 = (l+p)^2 = m_c^2 + l^2 + 2p\cdot l \approx l^2 + 2 l_0 m + m_c^2\;,
\end{equation}
which results in:
\begin{equation}
|\Gamma_0|^2 = 1 + \frac{ l_0}{2m}\;.
\end{equation}
If $|\Gamma_0|$ is going to be zero it requires $l_0 = -2m$ which means that the charm quark
gets a negative energy, which is unphysical. Now we can investigate how much the incoming
charm quark needs to be off-shell so that we can work with a real gluon field instead:
\begin{equation}
m_c^2 = (l+p) = (1-\epsilon)m_c^2 +   2p\cdot l  \rightarrow \epsilon \approx \frac{2l_0}{m}\;.
\end{equation}
Then, for small enough $l_0$,
we can again treat the off-shellness as a velocity correction along the many others 
we have neglected in the calculation. 
\newpage
\end{appendix}
\end{document}